\documentclass{aa}
\usepackage{amsmath}
\usepackage{txfonts}
\usepackage{graphicx}
\usepackage{natbib}
\bibpunct{(}{)}{;}{a}{}{,} 

\def \j {\sigma}
\def \Frac {\displaystyle\frac}
\def \be {\begin{equation}}
\def \ee {\end{equation}}
\def \w {8.8cm}
\def \lw {18cm}
\def \hd {HD\,202206}
\def \mjup {\,M_\mathrm{jup}}
\def \sb {b}
\def \sc {c}
\def \rsc {\sqrt{\chi_\mathrm{r}^2}}

\def \f1 {
\begin{figure}[ht]
   \centering
    \includegraphics[height=\w,angle=270]{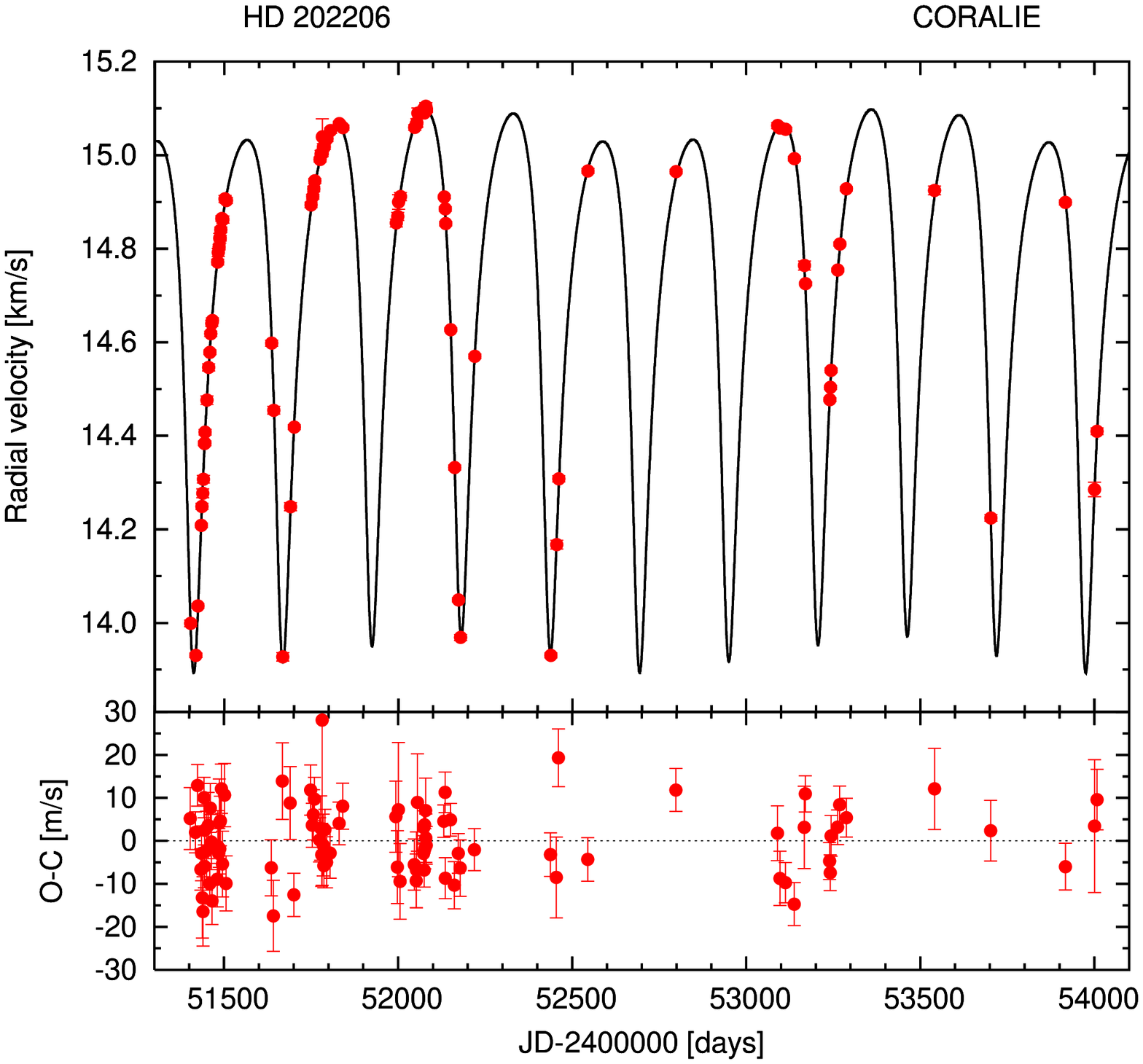}
  \caption{CORALIE radial velocities for \hd\ 
    superimposed on a 3-body Newtonian orbital solution (Table\,\ref{tab:best_fit_cop_90}).
    \label{F1} }
\end{figure}
}

\def \repere {
\begin{figure}[ht]
    \centering
    \includegraphics[width=\w]{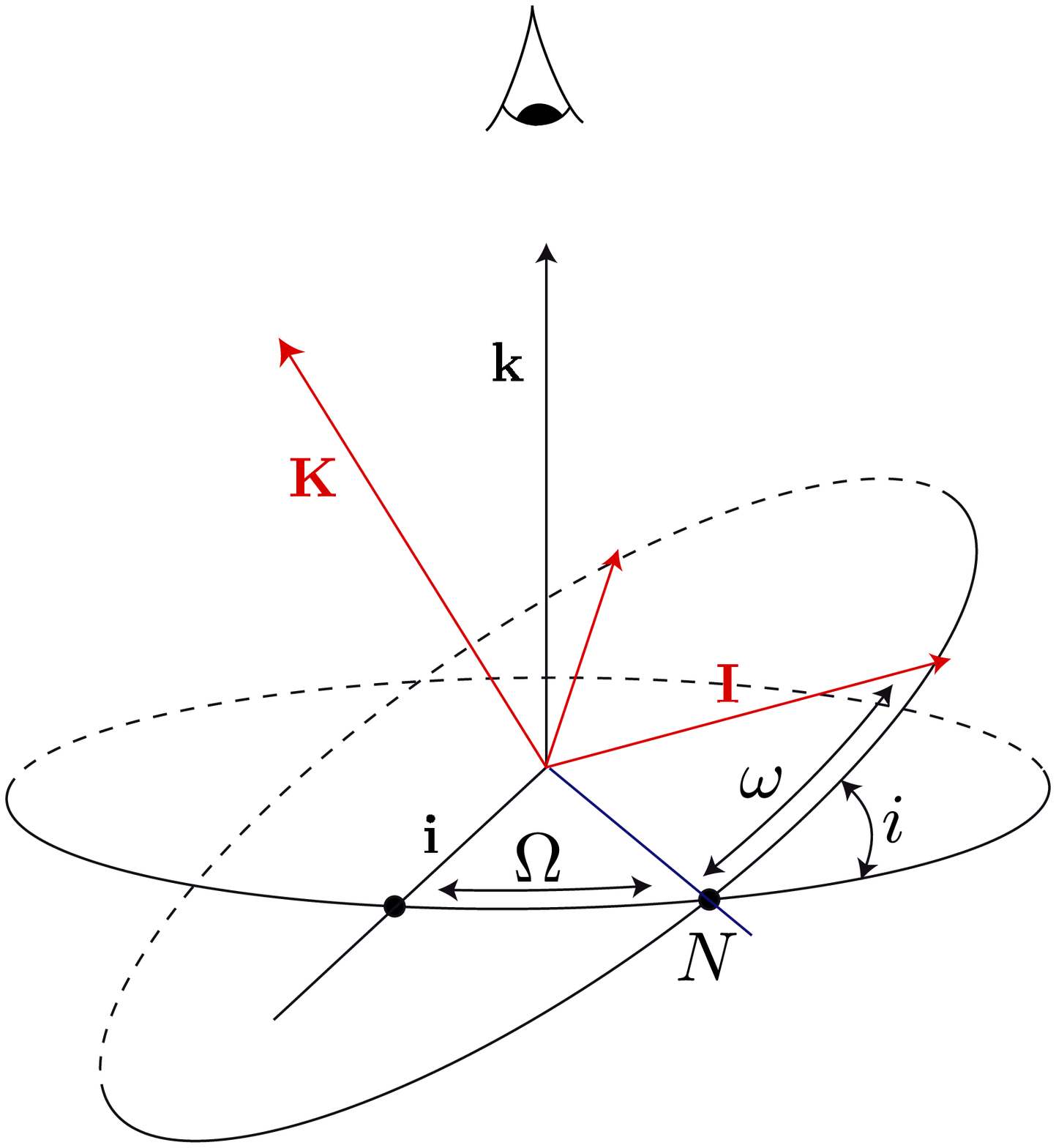}
    \caption{\label{fig:geometry} Angles defining the orbit's orientation in
     space. We follow the observers convention which sets the plane of sky as
     the reference frame, for which the edge-on coplanar configuration is
     $i_\sb=i_\sc=90\degr$ and $\Delta\Omega = 0\degr$. The
     observer is limited to the velocity projected on the $k$ axis.} 
\end{figure}
}

\def \bin {
\begin{figure}[ht]
    \centering
    \includegraphics[width=\w]{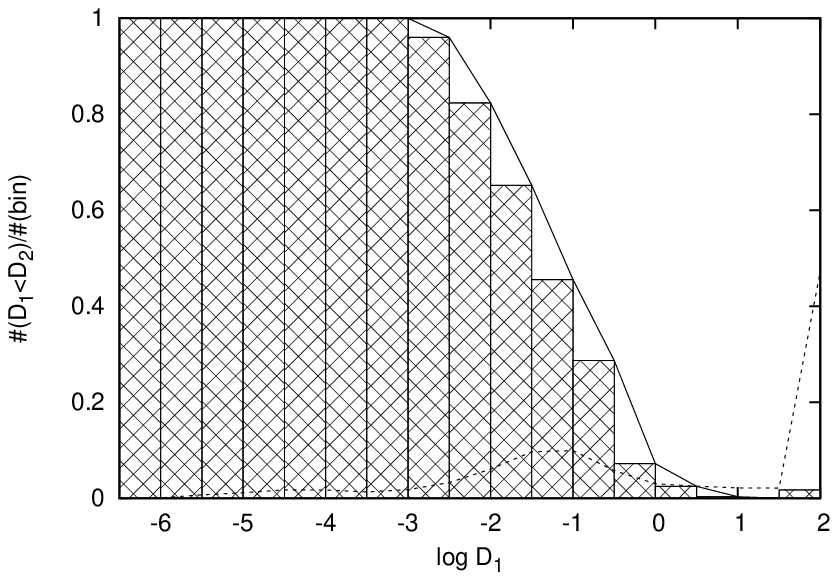}
    \caption{Distribution of $D_1 < D_2$ trajectories from the top panel of
Fig.~\ref{fig:aw_ae_90_cop}.
    Each trajectory integrated is binned with respect to its diffusion index
    $\log{D_1}$, after the diffusion grid has been smoothed.
    For each bin we compute the proportion of trajectories with a decreasing
    diffusion index over time: $D_1 < D_2$.
    The histogram shows the results for 0.5 wide bins. As expected from equation
    \ref{eq:dlim}, orbits with a high diffusion index $D_1$, and $D_1 < D_2$ are
    nearly inexistant ($\log{D_1} > -1$), while we observe the opposite
    situation for low diffusion index orbits ($\log{D_1} < -3$).}
    \label{fig:bin}
\end{figure}
}

\def \testdlim {
\begin{figure}[ht]
  \centering
  \includegraphics[width=\w]{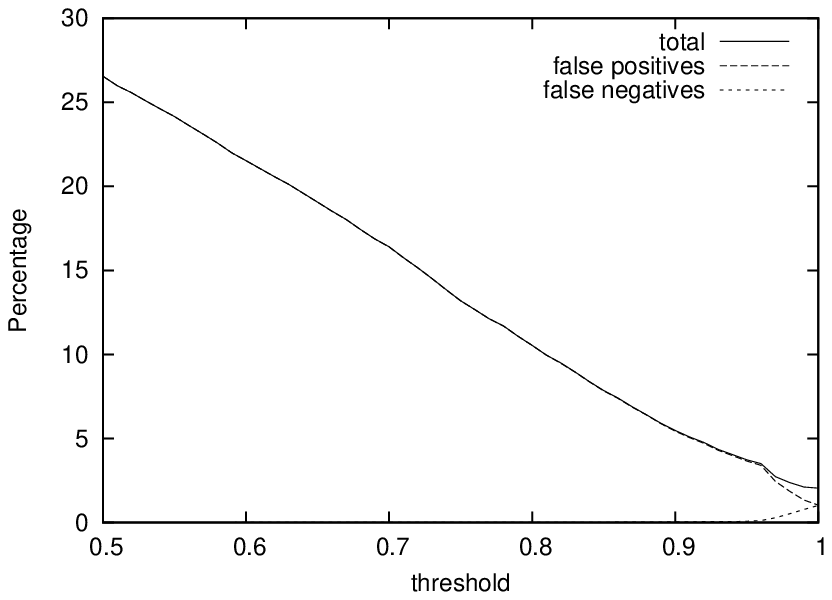}
  \caption{Percentage of orbits wrongly flagged stable (false stable) or
        unstable (false unstable). Once $D_\mathrm{lim}$ is choosen using a
        given percentage threshold (see figure \ref{fig:bin}), we have a
        criterion for stable ($D_1 < D_\mathrm{lim}$) and unstable orbits ($D_1
        > D_\mathrm{lim}$). We compare the results with a diffusion grid
        computed over a longer time interval (in this case $2\times 40000$
        years) taken as a reference. A few orbits deemed stable from the
        reference diffusion grid were thought to be unstable and vice versa. The
        solid line traces the number of those faulty orbits for different values
        of the threshold. It appears to be minimum between $0.95$ and $1$. The
        dotted line traces the false positives, and the dashed line, the false
        negatives (the solid one is the sum of those two).}
  \label{fig:testdlim}
\end{figure}
}

\def \bestfitcop {
\begin{figure}[ht]
  \centering
    \includegraphics{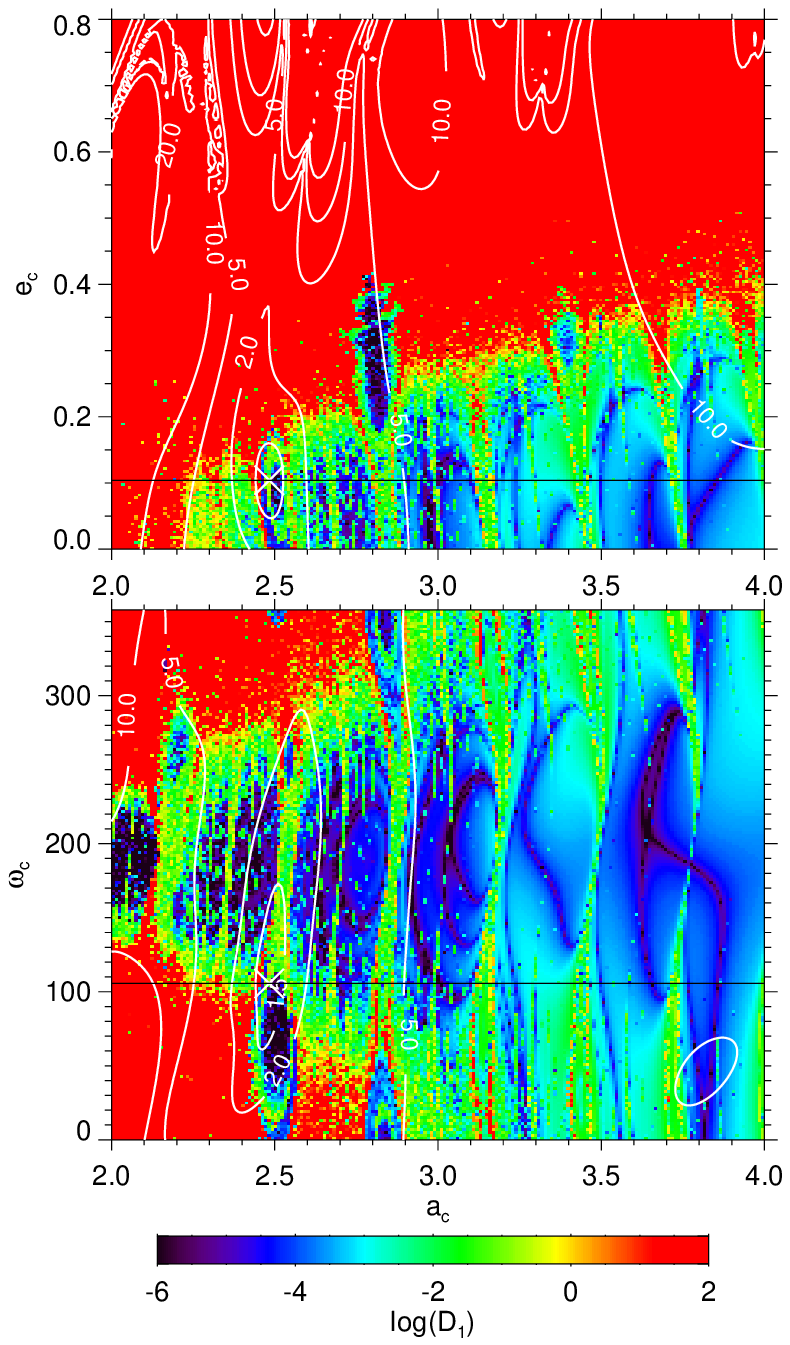}
    \caption{\label{fig:aw_90_cop_large} Global view of the dynamics of HD202206
         for variation of the semi-major axis and periastrum of the outer planet
         (bottom panel) or semi-major axis and eccentricity (top panel). The
         step-sizes for $a_\sc$, $\omega_\sc$, $e_\sc$ are respectively $0.005 AU$,
         $2\degr$ and $0.004$. The other parameters were
         kept constant and taken from
         the fit \textbf{S1} (Table \ref{tab:best_fit_cop_90}). The
               color
         scale is the stability index $\log{D_1}$ obtained
         through a frequency analysis of the longitude of the outer planet over two
         consecutive time intervals of $8000$ years. The level curves give the
         $\rsc$ value computed for each choice of parameters. The two horizontal black
         lines mark the intersection of the two grids.
         Most of the orbits are chaotic
               (yellow to red
         dots), but regular orbits zones exist
               (blue
         dots). One of these regular orbits regions lies inside the low
         $\rsc$ region: the dark area around $a_\sc=2.5$ AU and $\omega_\sc=50\degr$.
         It corresponds to the stable island of the $1/5$ mean motion resonance. The
         cross marks the \textbf{S1} orbital solution.}
\end{figure}
}

\def \awcoplarge {
\begin{figure}[ht]
  \centering
    \includegraphics{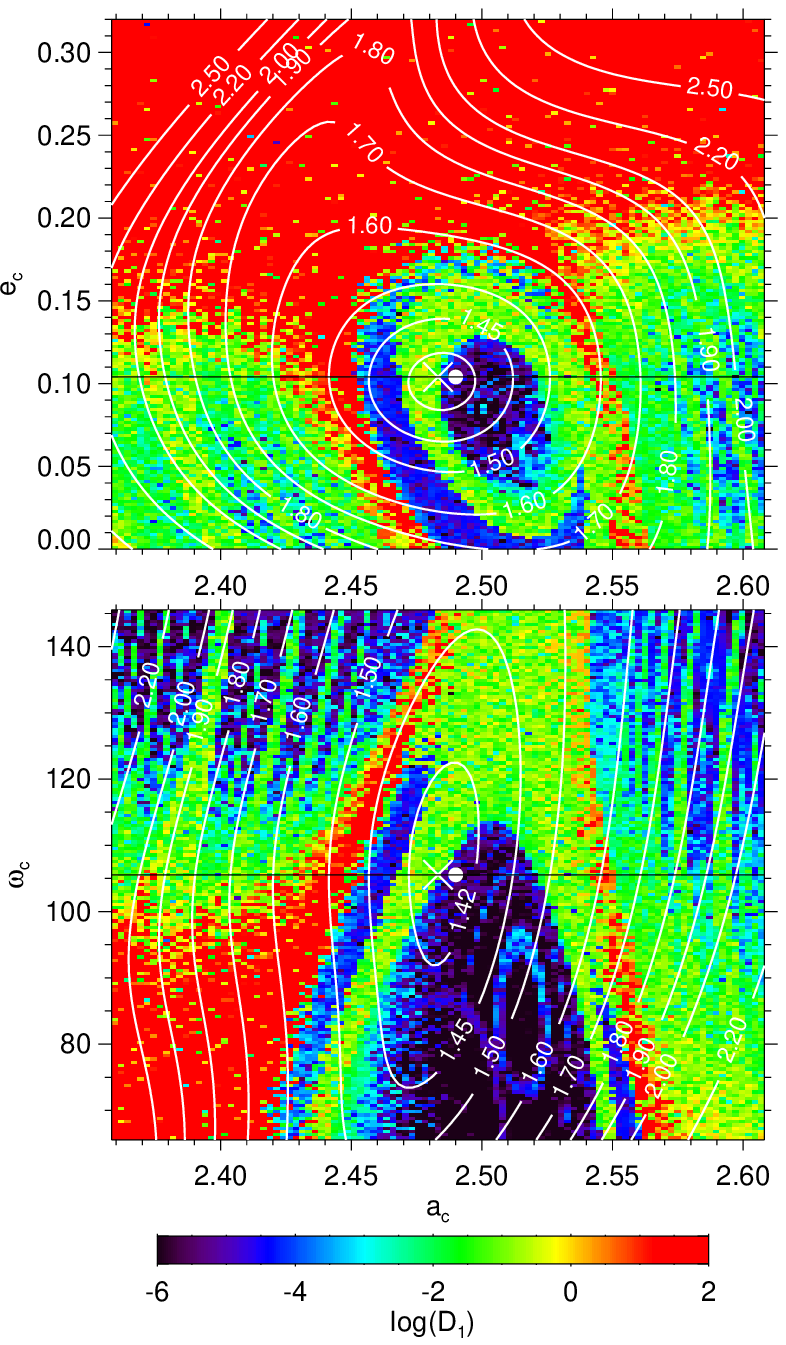}
    \caption{\label{fig:aw_ae_90_cop} Global view of the dynamics of HD202206
         for variation of the semi-major axis and periastrum (bottom), and semi-major
         axis and eccentricity (top) of the outer planet. The step-sizes are
         respectively $0.0025$ AU, $0.5\degr$, and $0.002$ for the eccentricity. The
            color
         scale is the stability index ($\log{D_1}$), and the level curves give the $\rsc$
         values (see Fig.~\ref{fig:aw_90_cop_large}). The cross marks the best fit
         \textbf{S1}, and the horizontal line in each panel, the orbits common to both
         maps (ie. the intersection of each map). The best fit lies very close to a stable
	 resonant island where we can pick a stable solution with a $\chi^2$ only marginally
	 higher than that of the best fit. As an example we picked such a solution (\textbf{S2}),
	 marked by a white filled circle, by slightly increasing $a_c$ from 2.4832 AU to 2.49 AU.
	 The complete set of orbital elements for \textbf{S2} is given in Tab.\ref{tab:90_stable}}
\end{figure}
}

\def \thetar {
\begin{figure}[ht]
 \includegraphics[width=\w]{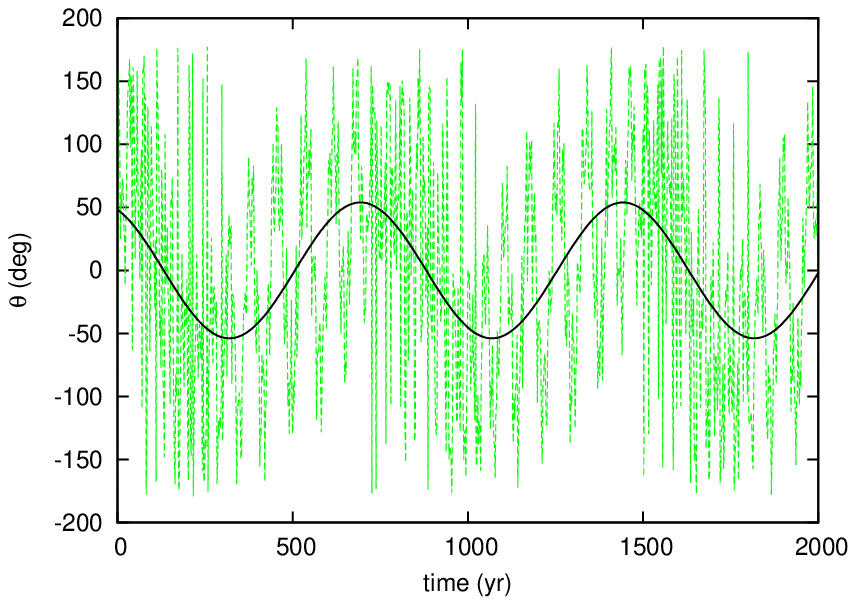}
 \caption{\label{fig:S90_thetaR_bw} Time variation of the resonant argument
        $\theta=\lambda_\sb - 5\lambda_\sc +  \varpi_\sb + 3\varpi_\sc$ for the orbital 
        solution \textbf{S2} (green line).
        $\theta$ is in libration around $\theta_0 = 0^\circ$,
	with a modulation of two terms of period $P_{\theta} = 80.13$ years,
	and $P_{\Delta\varpi}=749.10$ years,
        amplitudes of about $35$ and $50$ degrees respectively.
	The black line shows the first secular term contribution.}
\end{figure}
}

\def \testparts {
\begin{figure*}[ht]
 \includegraphics[width=\lw]{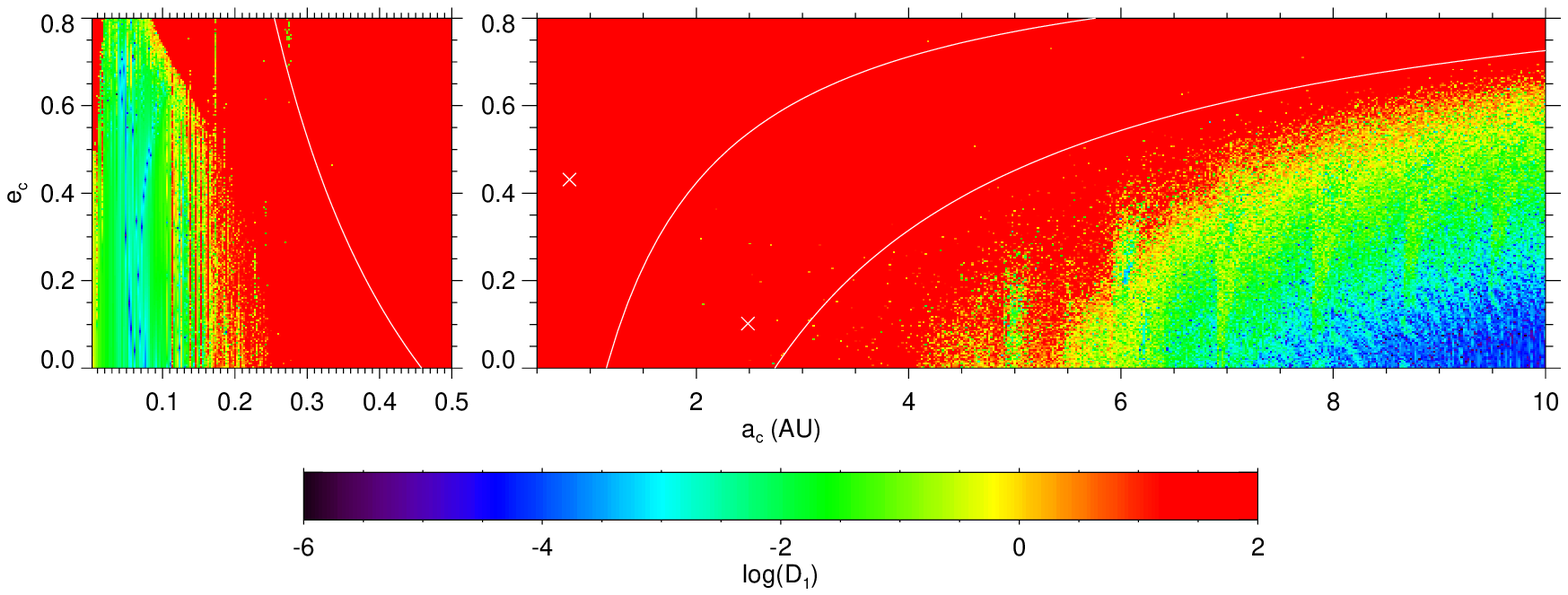}
 \caption{\label{fig:testpart} Mean motion diffusion of tests particles.
    integration S2 solution with massless particles over 16000yr. We compute two
determination $n$ and $n^\prime$ of each particules mean motion over two
consecutive time intervals of 8000 yr.
    The color scale code the stability index $\log{|n-n^\prime|}$.
    Initial conditions for the massless particles are
$(i_\mathrm{p},M_\mathrm{p},\omega_\mathrm{p},\Omega_\mathrm{p}) =
(90\degr,0\degr,0\degr,0\degr)$.
    On both panel we vary the initial eccentricity of the test particules from
    $0$ to $0.8$ with a step size of $0.004$.
    On the left panel the semi-major axis varies from $0.05$ AU to $0.5$ AU
    with a step size of $0.0025$ AU, and on the right panel it varies from
    $0.5$ AU to $10$ AU with a step size of $0.05$.
    The white crosses mark the position of the two planets, and the collision
    lines are traced with white lines.
    }
\end{figure*}
}

\def \torus {
\begin{figure}
\centering
\includegraphics[width=6cm]{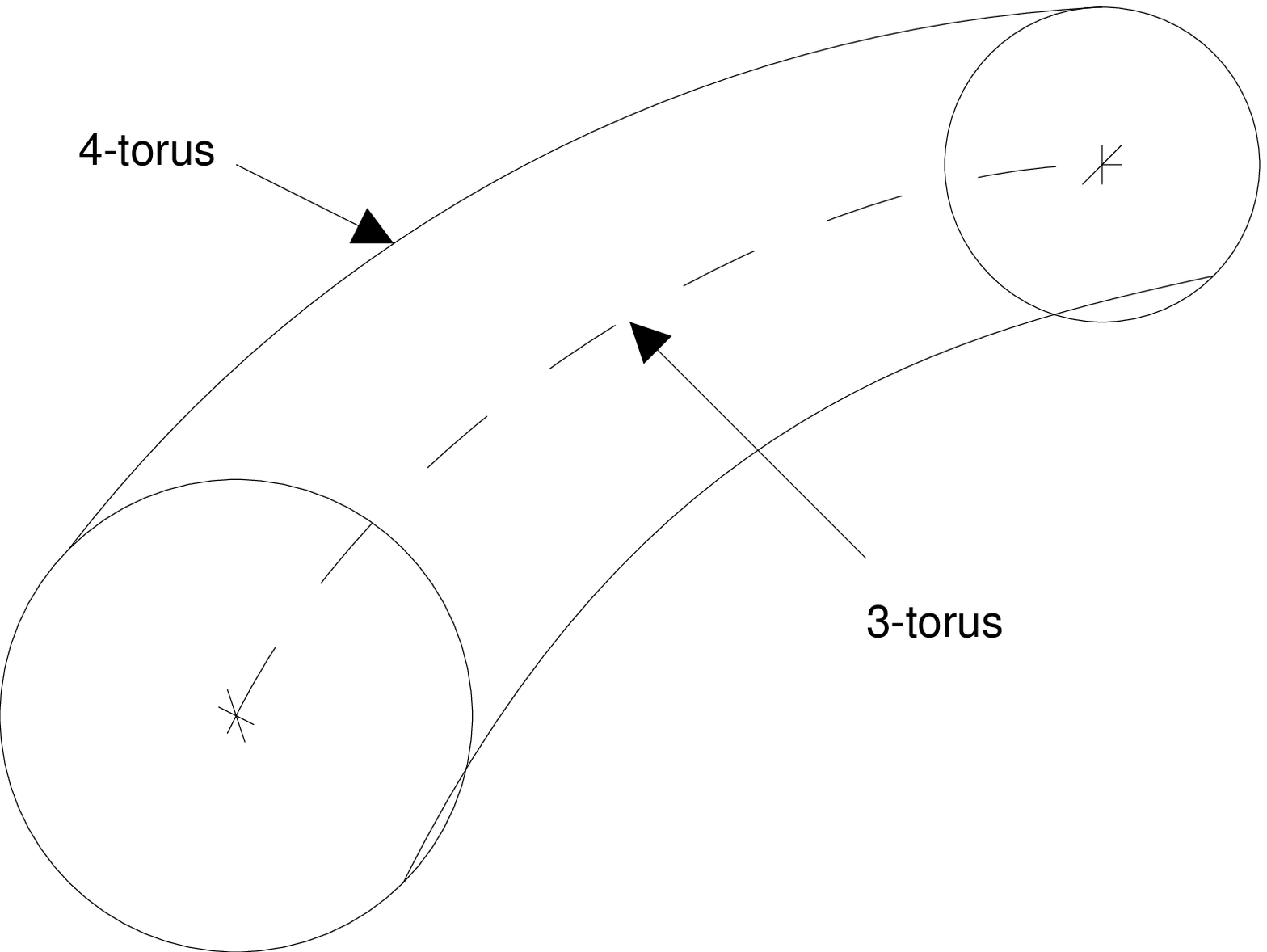}
\caption{\label{fig:torus_schematic} Schematic representation of the 3-torus of the center of libration.
We represent the 4-torus of a given resonant orbit by a 2-torus (a doughnut) as
it is not possible to represent it otherwise.
The center of libration 3-torus is then represented by the circle in the center of the interior of the doughnut.}
\end{figure}
}

\def \tores1 {
\begin{figure}
\includegraphics{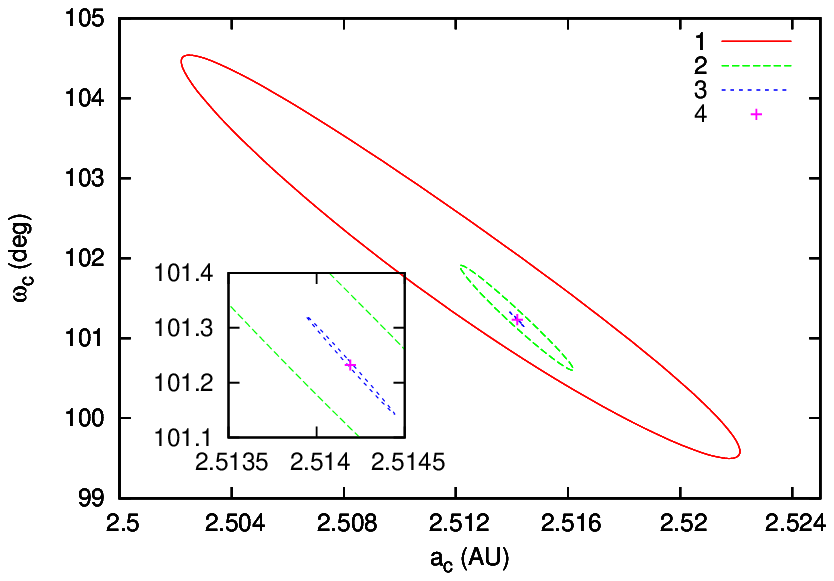}
\caption{\label{fig:torus_num} Projection of a section of the 4-torus of each
step's trajectory in the $(a_\sc,\omega_\sc)$ plane. Each torus is approximated
using a (truncated) quasi-periodic decomposition of the trajectory.}
\end{figure}
}

\def \steps {
\begin{figure}
\includegraphics{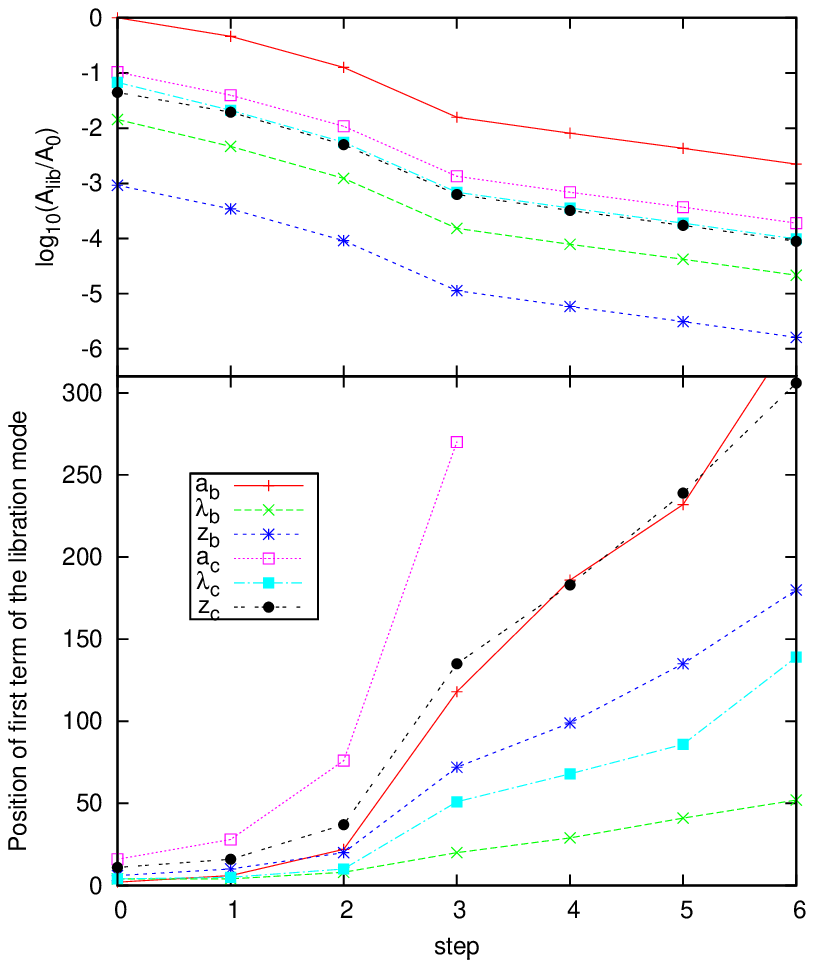}
\caption{\label{fig:steps} Evolution of the amplitude of the libration mode
in the quasi-periodic decomposition of each variable at each step.
In the top panel we plot the relative amplitude of the first term depending on
$f_0$ compared to the first not constant term.
In the bottom panel we plot the position of this first term in the decomposition.
The first step (abscissa 0) is the \textbf{S2} orbital solution, and the last step (abscissa 6) is the orbital solution \textbf{S3} (table \ref{tab:s3}).
}
\end{figure}
}

\def \aelarge {
    \begin{figure*}[ht]
     \includegraphics{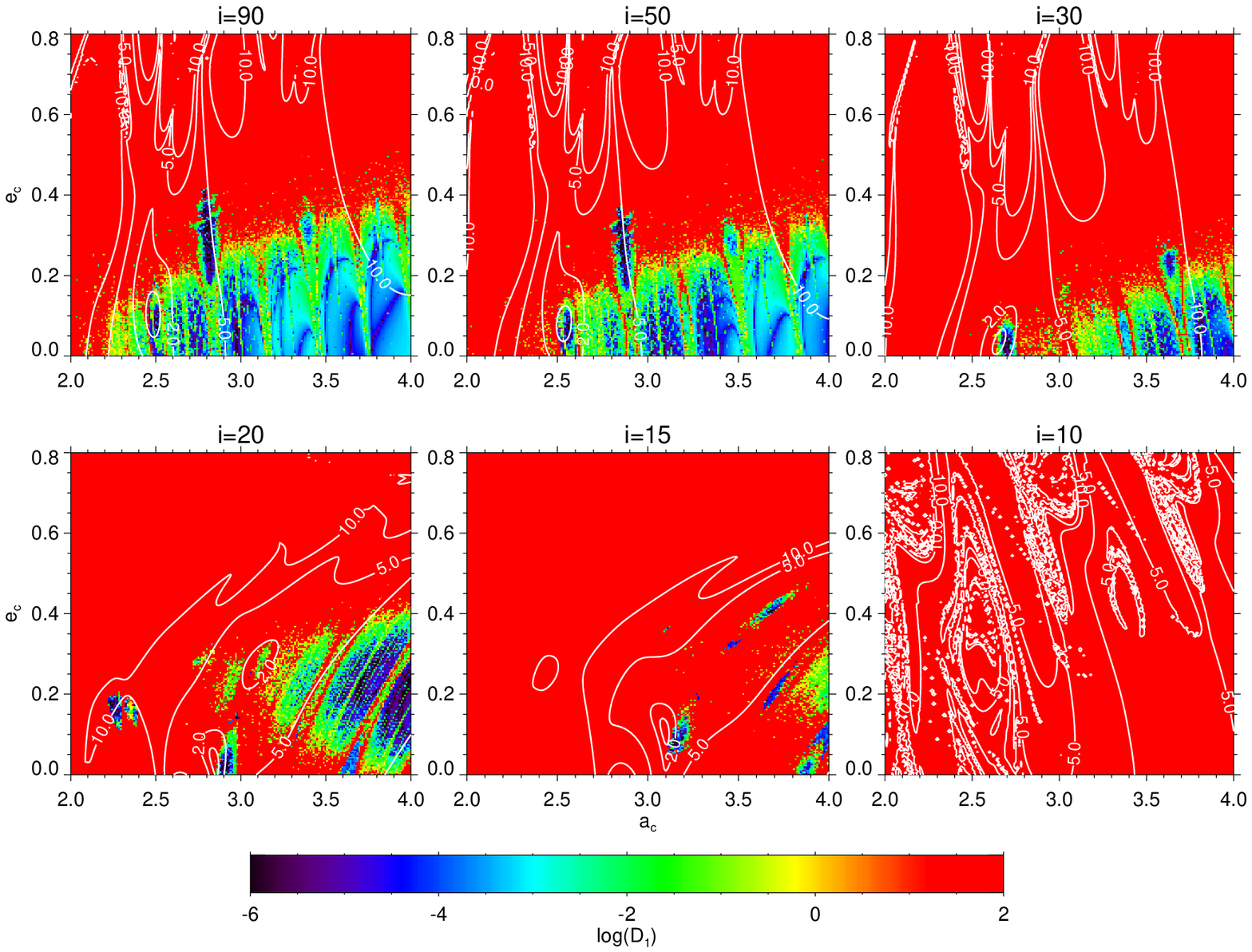}
     \caption{\label{fig:i_ae_large} Dynamics of a coplanar HD202206 system for
different value of the inclination $i$. Each panel is a diffusion map in the
$(a_\sc,e_\sc)$ plane of initial conditions constructed the same way as the
top panel of Fig.~\ref{fig:aw_90_cop_large} (the top-left panel is in fact a
copy). For each value of inclination $i$, a new fit, assuming a coplanar system,
is computed.}
    \end{figure*}
}

\def \progradebest {
\begin{figure}[ht]
 \includegraphics[width=\w]{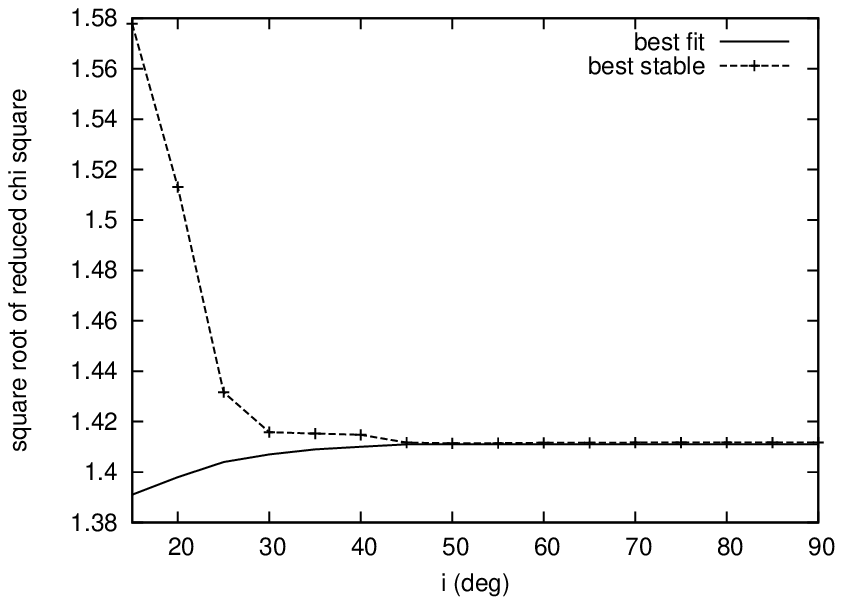}
 \caption{\label{fig:cop_best} Evolution of the best fit $\rsc$ (solid curve)
and an estimation of the best stable orbits $\rsc$ in function of the
inclination $i$. For each value of $i$ between $90\degr$ and $10\degr$, a new
fit, assuming a coplanar configuration, is computed. We then look for stable
orbits around each of these orbital solutions in a $(a_\sc,e_\sc)$ plane of
initial conditions.}
\end{figure}
}

\def \progradepct {
\begin{figure}[ht]
 \includegraphics{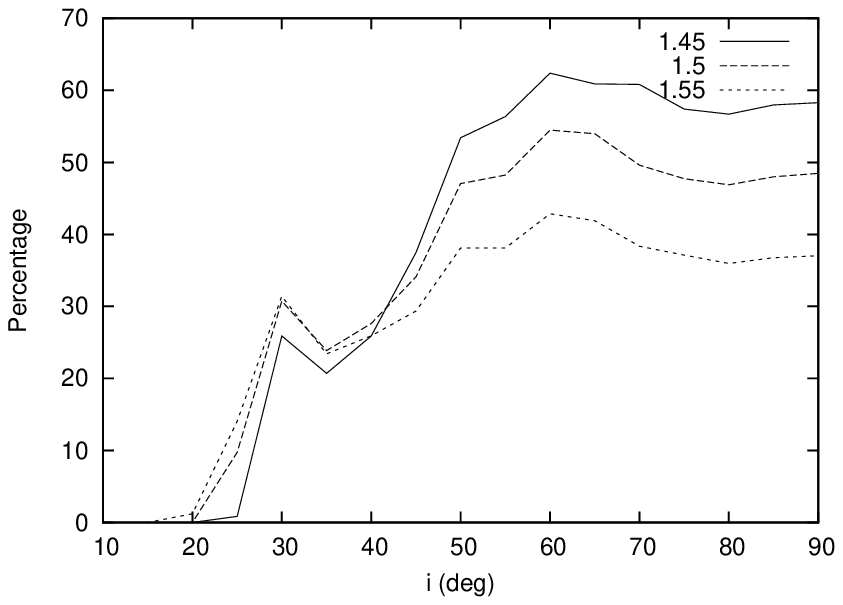}
 \caption{\label{fig:cop_pct} Percentage of stable orbits inside a given $\rsc$
level curve against $i$ (see Fig.~\ref{fig:cop_best}).}
\end{figure}
}

\def \deltavr {
\begin{figure}
 \includegraphics{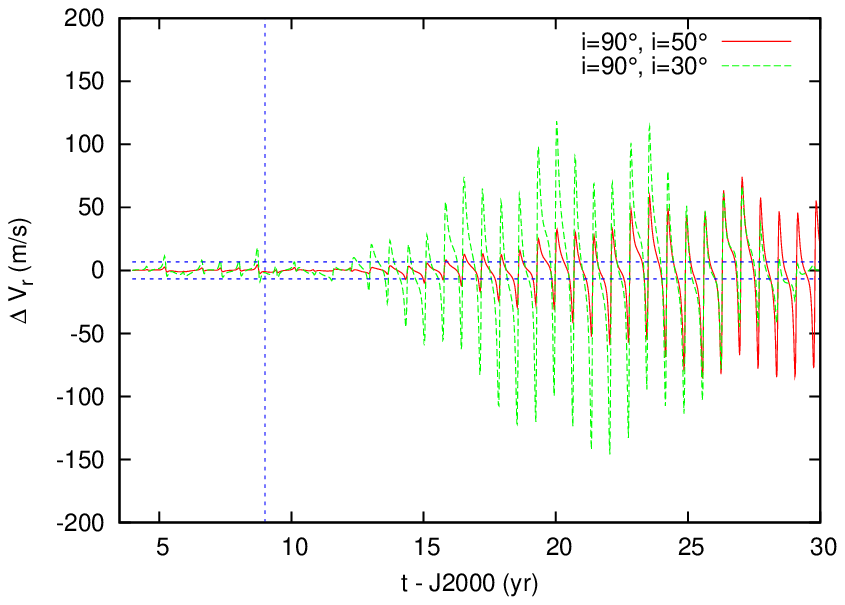}
 \caption{\label{fig:Delta_vr} Differences between the radial velocity of a
        stable solution with $i=90\degr$, and $i=50\degr$ (red curve), or
        $i=30\degr$ (green curve). The starting epoch of the three integrations
        is $2453000.00 JD$. The blue horizontal lines represents the precision
        of the current set of data, obtained with the CORALIE instrument. The
        vertical blue line marks the beginning of 2009.}
\end{figure}
}

\def \freqf {
\begin{figure}[ht]
 \includegraphics[width=\w]{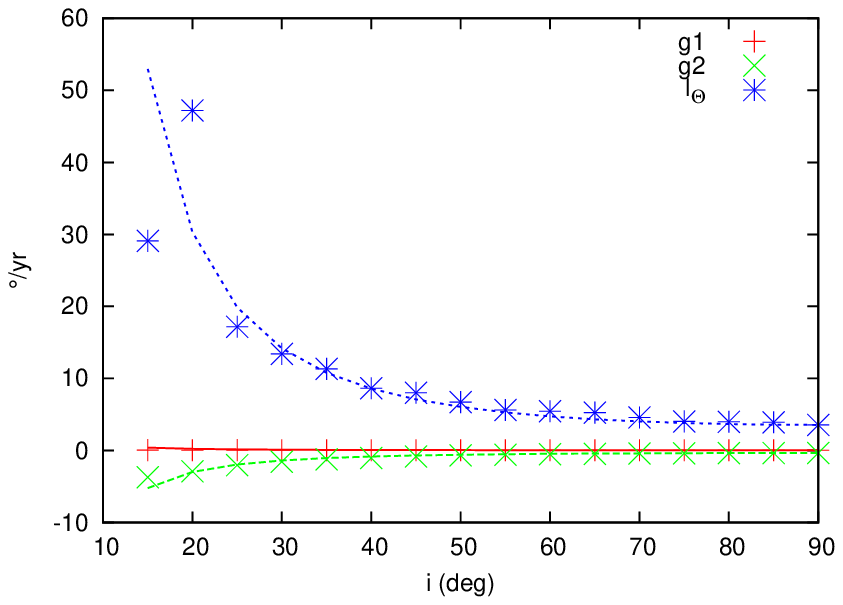}
 \caption{\label{fig:freqF}Libration and secular frequencies. For each inclination $i$ we pick a
    stable orbit with a low $\chi^2$ and integrate it over 1 million
    years. Using frequency analysis, we determine for each orbit its
    fundamental frequencies. The solid curves represent, for each frequency, a
    $1/\sin^2{i}$ law.}
\end{figure}
}

\def \mutone {
\begin{figure}[ht]
  \includegraphics[width=\w]{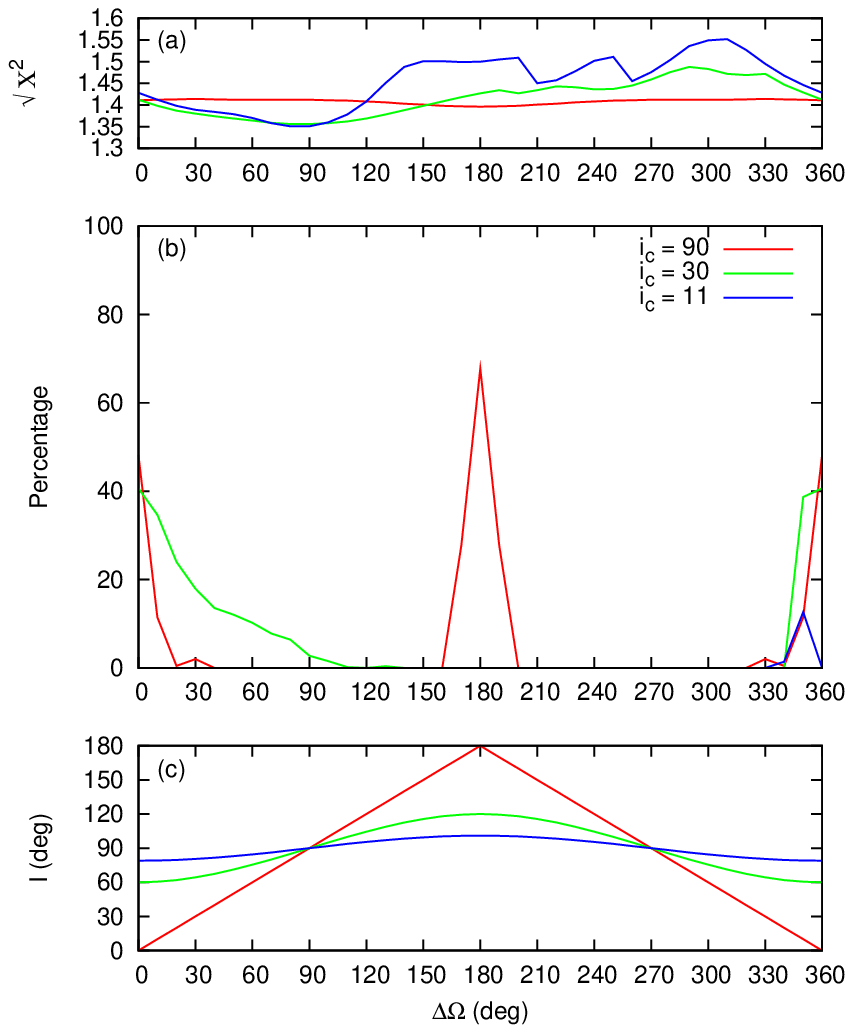}
  \caption{\label{fig:mutone} Stable configurations for $i_b = 90\degr$, that is for the minimum mass
  of planet $b$. For each value of $î_c$ in $(90\degr, 30\degr, 11\degr)$, and $\Delta\Omega$ between
  $0\degr$ and $360\degr$ we compute diffusion index over grids in the $(a_c, e_c)$ plane of initial
  conditions with step sizes of respectively 0.0025 AU and 0.004. Each grid is also centered on the
  minimum $\chi^2$ computed for each couple $(i_c, \Delta\Omega)$.
  We plot against $\Delta\Omega$ the square root of the minimum $\chi^2$ in the top panel, and the
  proportion of stable orbits inside $\rsc = 1.5$ in the middle panel.
  The mutual inclination corresponding to each triplet $(i_b, i_c, \Delta\Omega)$ is plotted in the
  bottom panel.
  }
\end{figure}
}

\def \muttwo {
\begin{figure}[ht]
  \includegraphics[width=\w]{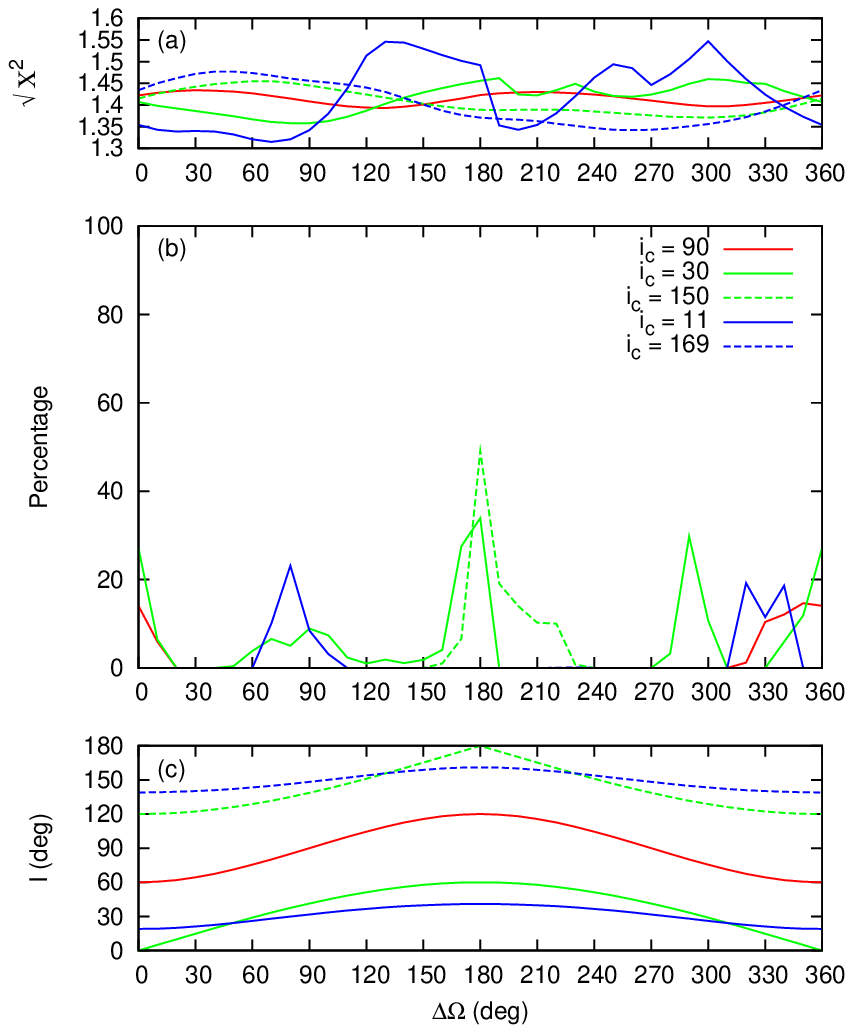}
  \caption{\label{fig:muttwo} Stable configurations for $i_b = 30\degr$, that is for approximately
  twice the minimum mass: $m_b \simeq 33.5 \,M_{\mathrm{jup}}$.
  See Fig.~\ref{fig:mutone} caption for more details.
  }
\end{figure}
}

\begin{document}
\title{Dynamical stability analysis of the HD202206 system and
    constraints to the planetary orbits\thanks{Based
    on observations made with the CORALIE instrument on the EULER 1.2m
    telescope at La Silla Observatory under the ?? programme ID
    ??. The table with the radial velocities is available in 
    electronic form at the CDS via anonymous ftp to cdsarc.u-strasbg.fr
    (130.79.128.5) or via http://cdsweb.u-strasbg.fr/cgi-bin/qcat?J/A+A/????  
    }  }

\author{J.~Couetdic\inst{1}
    \and J.~Laskar\inst{1}
    \and A.C.M.~Correia\inst{1,2}
    \and M.~Mayor\inst{3}
    \and S.~Udry\inst{3}}

\institute{Astronomie et Syst\`{e}mes Dynamiques, IMCCE-CNRS UMR 8028,
Observatoire de Paris, UPMC, 77 avenue
    Denfert-Rochereau, 75014 Paris, France
    \and Departamento de F\'{\i}sica, Universidade de Aveiro, Campus
    de Santiago, 3810-193 Aveiro, Portugal
    \and Observatoire de Gen\`eve, 51 ch. des Maillettes, 1290 Sauverny,
    Switzerland}

\abstract
    {
	Long-term precise Doppler measurements with the CORALIE spectrograph
	revealed the presence of two massive companions to the solar-type
	star HD202206.
	Although the three-body fit of the system is unstable, it was shown
	that a 5:1 mean motion resonance exists close to the best fit,
        where the system is stable.
 	It was also hinted that stable solutions with a wide range of mutual
	inclinations and low O-C were possible.
    }
    {
	We present here an extensive dynamical study of the HD202206 system
	aiming at constraining the inclinations of the two known companions, 
	from which we derive possible ranges of value for the companion masses.
    }
    {
	We consider each inclination and one of the longitude of ascending node
	as free parameters.
	For any chosen triplet of these parameters, we compute a new fit.
	Then we study the long term stability in a small (in terms of
	O-C) neighborhood using Laskar's frequency map analysis.
	We also introduce a numerical method based on frequency analysis to
	determine the center of	libration mode inside a mean motion resonance.
    }
    {
        We find that acceptable coplanar configurations (with low $\chi^2$
	stable orbits) are limited to inclinations to the line of sight between
	$30\degr$ and $90\degr$.
	This limits the masses of both companions to roughly twice the minimum:
	$m_b \in [16.6\,M_{Jup};\,33.5\,M_{Jup}]$ and $m_c \in
	[2.2\,M_{Jup};\,4.4\,M_{Jup}]$.
	Non coplanar configurations are possible for a wide range of mutual
	inclinations from $0\degr$ to $90\degr$, although $\Delta\Omega = 0
	[\pi]$ configurations seem to be favored.
	We also confirm the 5:1 mean motion resonance to be most likely.
	In the coplanar edge-on case, we provide a very good stable solution in
	the resonance, whose $\chi^2$ does not differ significantly from the
	best fit.
	Using our method to determine the center of libration, we further refine
	this solution to obtain an orbit with a very low amplitude of libration,
	as we expect dissipative effects to have dampened the libration.
    }
    {}


\maketitle

\section{Introduction}
    The CORALIE planet-search program in the southern hemisphere has found two
    companions around the HD202206 star.
    The first one is a very massive body with $17.5\mjup$ minimum mass
    \citep{Udry_etal_2002}, while the second companion is a $2.4\mjup$ minimum
    mass planet \citep{Correia_etal_2005}. The parent star has a mass of
    $1.044$ solar masses, and is located $46.3$ pc from the Solar System.
    The HD202206 planetary system is an interesting case to investigate the brown
    dwarf desert since the more massive companion  can be either a huge planet
    (formed in the circumstellar disk) or a low-mass brown dwarf candidate.
    
    \citet{Correia_etal_2005} found that the orbital parameters obtained with
    best fit for the two planets leads to a catastrophic events in a short time 
    (two keplerians fit and full three-body fit alike).
    This was not completely unexpected given the very large eccentricities
    ($0.435$ and $0.267$) and masses of the two planets.
    Using frequency analysis \citep{Laskar_1990,Laskar_1993,Laskar_1999} they
    performed a study on the global dynamics around the best fit, and found that
    the strong gravitational interactions with the first companion made the
    second planet evolution very chaotic, except for initial conditions in the
    5:1 mean motion resonance.
    Since the associated resonant island actually lies close to the $\chi^2$
    minimum value of the best fit, they concluded that the system should be
    locked in this 5:1 resonance.
    Later on, Gozdziewski and co-workers also looked for stable solutions in this
    system using their GAMP algorithm \citep{Godziewski_etal_2006}.
    They provided two possible solutions (among many others), one coplanar, and 
    one with a very high mutual inclination.
    
    Since then, new data have been acquired using the CORALIE spectrograph. A new
    reduction of the data changed some parameters, including the mass of the
    HD202206 star. A new fit assuming a coplanar edge-on configuration was derived 
    from the new set of radial velocity data. This new solution still appears to
    be in the 5:1 mean motion resonance, but is also still unstable. The most striking
    difference from \citet{Correia_etal_2005} is the smaller eccentricity of HD202206\,c.
    
    In the present work, we will continue in more detail the dynamical study
    started in \citet{Correia_etal_2005}, using the new fit as a starting point.
    We also aim to find constraints on the
    orbital parameters of the two known bodies of this system, in
    particular the inclinations (and thus the real masses of the planets).
    \citet{Godziewski_etal_2006} already showed that stable fits could be
    obtained with different inclinations, using a particular fitting
    genetic algorithm that adds stability computation to select its
    populations (GAMP).
    Although very effective in finding a stable fit, this algorithm cannot
    find all possible solutions. We prefer here an approach which separates the
    fitting procedure from dynamical considerations, as it allows for a better
    assessment of the goodness of the fit, and whether the model is a good
    description of the available data. The trade-off is a more difficult
    handling of the high number of parameters.
      
    
    We briefly present the new set of data in Sect.~\ref{sec:data}, the
    numerical methodology in Sect.~\ref{sec:method}. We review in details the
    dynamics in the coplanar edge-on case in Sect.~\ref{sec:cop_edge}. We then
    release the constraint on the inclination of the system from the line of
    sight in Sect.~\ref{sec:cop}, and finally briefly investigate the
    mutually inclined configurations in Sect.~\ref{sec:mutual}.

\section{\label{sec:data}New orbital solution for the HD202206 system}
	\begin{table}[ht]
	\centering
	\caption{\label{tab:best_fit_cop_90} Best Newtonian fit \textbf{S1} for
	the HD202206 system assuming $i_\sb=i_\sc=90\degr$, and
	$\Delta\Omega=0\degr$.}
	\begin{tabular}{llcc}
	  \hline\hline
	    \textbf{Param.} & \textbf{S1} & \textbf{inner} & \textbf{outer} \\
	  \hline
	    $V_0$ & [km/s] & \multicolumn{2}{c}{$14.729 \pm 0.001$} \\
	    $P$ & [days] &  $256.389 \pm 0.044$ & $1397.445 \pm 19.056$ \\
	    $K$ & [m/s] &  $564.82 \pm 1.42$ & $38.08 \pm 1.21$ \\
	    $e$ & & $0.431 \pm 0.001$ & $0.104 \pm 0.024$ \\
	    $\omega$ & [deg] & $161.91 \pm 0.27$ & $105.56 \pm 15.77$ \\
	    $M+\omega$ & [deg] & $239.016 \pm 0.13$ & $250.38 \pm 2.71$ \\
	  \hline
	    $a$ & [AU] & 0.8053 & 2.4832 \\
	    $i$ & [deg] & 90 & 90 \\
	    $\Omega$ & [deg] & 0 & 0 \\
	    $m$ & [$\mbox{M}_{\mbox{\begin{tiny}Jup\end{tiny}}}$] & 16.59 & 2.179 \\
	  \hline
	    Date & [JD-2400000] & \multicolumn{2}{c}{53000.00} \\
	    rms  & [m/s] & \multicolumn{2}{c}{7.4544} \\
	    $\rsc$ & & \multicolumn{2}{c}{1.411} \\
	  \hline
	\end{tabular}
	Errors are given by the standard deviation $ \sigma $. This fit
	corresponds to a coplanar system seen edge-on, with minimum values for the
	masses. 
	\end{table}
    
    The CORALIE observations of HD202206 started in August 1999 and the last point
    acquired in our sample dates from September 2006, corresponding to about seven
    years of observations and 92 radial-velocity measurements.
    Using the iterative Levenberg-Marquardt method \citep{Press_etal_1992}, 
    we fit the observational data using a 3-body Newtonian model,
    assuming co-planar motion perpendicular to the plane of the sky, similarly to
    what has been done in \citep{Correia_etal_2005,Correia_etal_2009}.
    Notice that we changed the reference date with respect to the solution in
    \citet{Correia_etal_2005}.
    The mass of the star has also been updated to $1.044\,M_\odot$
    \citep{Sousa_etal_2008}.
    This fit yields two planets with an adjustment of $\rsc=1.41$ and
    $rms=7.45\,\mathrm{m s}^{-1}$, slightly above the photon noise of the instrument
    which is around $6.69 \,\mathrm{ms}^{-1}$.
    We confirm the already detected planets \citep{Udry_etal_2002,Correia_etal_2005}
    with improved orbital parameters, 
    one at $P=254.8$\,day, $e=0.431$, and a minimum mass of
    $16.6\,M_\mathrm{Jup}$, and the other at $P=1397$\,day, $e=0.104$, and a minimum 
    mass of $2.18\,M_\mathrm{Jup}$ (Table\,\ref{tab:best_fit_cop_90}).
    In figure~\ref{F1} we plot the observational data superimposed on the best
    fitted solution.
    
        \f1
    
    We also fitted the data with a 3-body Newtonian model for which
    the inclination of the orbital planes, as well as
    the node of the outer planet orbit, were free to vary. 
    We were able to find a wide variety of configurations, some with low inclination values
    for one or both planets, that slightly improved our fit to a minimum $
    \rsc = 1.30 $ and $rms$\,=\,$7.08\,\mathrm{m/s}$. 
    However, all of these determinations remain uncertain, and since we also increase
    the number of free parameters by three, we cannot say that there has been an
    improvement with respect to the solution presented in Table\,\ref{tab:best_fit_cop_90}. 

\section{\label{sec:method}Numerical set-up}
  \subsection{\label{sec:conv}Conventions}
    In this paper, the subscripts $\sb$ and $\sc$ will respectively refer to the
    body with shortest and longest orbital period (inner and outer).
    The initial conditions ill-constrained by the radial velocity data are
    $i_\sb,i_\sc,\Omega_\sb,\Omega_\sc$ (the inclinations, and longitude of
    ascending nodes).
    We are using the observers convention which sets the plane of sky as the
    reference plane (see fig \ref{fig:geometry}).
    As a consequence, the nodal line is in the plane of sky, and has no
    cinematic impact on the radial velocities.
    From the dynamical point of view, only the difference between the two lines 
    of nodes $\Delta\Omega$ matters.
    In particular the mutual inclination depends on this quantity, through:
	
	\be
	    \label{eq:mutual_inclination}
	  \cos{I} = \cos{i_\sb}\cos{i_\sc} + \sin{i_\sb}\sin{i_\sc}\cos{\Delta\Omega}\ .
	\ee
	
	\repere
	
    $\Omega_\sb$ can thus always be set to $0\degr$ in the initial conditions,
    which leads to $\Delta\Omega = \Omega_\sc$, and only three parameters are
    left free $(i_\sb,i_\sc,\Omega_\sc)$.
    They are connected to the three interesting unknowns of the system,
    namely the mutual inclination $I$ (Eq.\,\ref{eq:mutual_inclination}), and
    the two planetary masses $m_\sb$, $m_\sc$ as: 
    
	\be
	    \label{eq:m1}
	    \left\{
	    \begin{aligned}
	    \Frac{m_\sb (m_\ast + m_\sb)^{1/3}}{m_\ast + m_\sb+m_\sc} =&
		\Frac{K_\sb P_\sb^{1/3}\sqrt{1-e_\sb^2}}{(2\pi G)^{1/3} \sin{i_\sb}} \ ,\\
	    \Frac{m_\sc (m_\ast + m_\sc)^{1/3}}{m_\ast + m_\sb+m_\sc} =&
		\Frac{K_\sc P_\sc^{1/3}\sqrt{1-e_\sc^2}}{(2\pi G)^{1/3} \sin{i_\sc}} \ ,
	    \end{aligned}
	    \right.
	\ee
	
    where $ m_\ast $ is the star mass,
    $ K_\j $ the amplitude of the radial velocity variations, $ P_\j $ the
    orbital period, $ e_\j $ the eccentricity, and $ G $ the gravitational
    constant.
    Basically, choosing the values of the inclinations is akin to setting the two
    companions masses. And for given values of $(i_\sb,i_\sc)$, we can control
    the mutual inclination with $\Omega_\sc$ through Eq.~\ref{eq:mutual_inclination}.
	
    Note that the denominator on the left hand sides in Eq.~\ref{eq:m1} is the
    total mass of the system. This term comes from the transformation to the
    barycentric coordinates system.
    As a consequence, the two equations are coupled.
    Of course, we can usually neglect the
    companion masses in this term, and decouple the equations.
    However when we change the inclinations,
    the planetary masses will grow to a point where this approximation
    is no longer valid.
    Here we always solve the complete equations, regardless of the inclinations.

    As long as the companions masses are small compared to the primary, they are
    scaled to $1/\sin{i}$ to a good approximation.
    We can define two factors $k_\sb = 1/\sin{i_\sb}$ and $k_\sc = 1/\sin{i_\sc}$.
    With $m_\sb^{(0)}$ and $m_\sc^{(0)}$ the minimum masses obtained for the edge-on
    coplanar case (see section \ref{sec:cop_edge} and table \ref{tab:best_fit_cop_90}),
    we can write:
	\begin{equation}
	    \label{eq:mfactor}
	    \begin{aligned}
	     m_\sb \approx & \, k_\sb m_\sb^{(0)} = \frac{1}{\sin{i_\sb}} m_\sb^{(0)} \ ,\\
	     m_\sc \approx & \, k_\sc m_\sc^{(0)} = \frac{1}{\sin{i_\sc}} m_\sc^{(0)} \ .
	    \end{aligned}
	\end{equation}
    For a given factor $k$, two values of the inclination are possible:
    $x$ and $\pi - x$ (where $x\in[0;\pi/2]$).
    For instance $i=30\degr$ and $i=150\degr$ give $k=2$.
    
    Additionally, for a given pair $(i_\sb,i_\sc)$, the accessible mutual
    inclinations $I$ are limited.
    Since the inclinations $i_\sb$ and $i_\sc$ are angles between
    $0$ and $\pi$ excluded, prograde coplanar configurations are only 
    possible for $i_\sb = i_\sc$ and $\Delta\Omega = 0$ (Eq.~\ref{eq:mutual_inclination}).
    Similarly, retrograde coplanar configurations are only possible for $i_\sb + i_\sc = \pi$
    and $\Delta\Omega = \pi$.
    This means that $k_\sb = k_\sc$ in both cases.
    More generally, $I \ge |i_\sb - i_\sc|$, and:
	\begin{equation}
	  \label{eq:mi}
	  \begin{aligned}
	    i_\sb + i_\sc \le \pi \ \Longrightarrow & \ I \le i_\sb + i_\sc \ , \\
	    i_\sb + i_\sc  >  \pi \ \Longrightarrow & \ I \le 2\pi - i_\sb - i_\sc \ .
	  \end{aligned}
	\end{equation}
    For any value of $I$, two values of $\Delta\Omega$ are possible: $x$
    and $-x$, since $\Delta\Omega$ appears through its cosine in Eq.~\ref{eq:mutual_inclination}.
    The extrema are obtained for $\Delta\Omega = 0$ (maximum), and $\Delta\Omega = \pi$
    (minimum)
    
    Finally we notice that one can restrict one of the two inclinations to the
    line of sight to $]0;\pi/2]$, which will be the case for $i_\sb$.

  \subsection{Fitting procedure}
    The influence of $i_b$, $i_c$, and $\Omega_c$ on the radial velocity data is 
    usually very small, and the perturbations depending on them have very long
    time scales.
    This makes any attempt to fit those parameters virtually impossible at present.
    Only with very strong mean motion resonances, such as observed in the GJ\,876
    system \citep{Laughlin_Chambers_2001, Correia_etal_2009}, can one hope to fit the inclinations.
    In this case the mean motion resonance introduces important short time scale
    terms (compared with the precision and time span of the observations).
    For the HD202206 system, the set of radial velocity data does not cover a 
    long enough period of time .

    Since the three parameters $i_\sb$,$i_\sc$ and $\Omega_\sc$ are very poorly
    constrained by the radial velocity data, we cannot fit the data with a model
    which includes them.
    Instead, for any chosen ($i_\sb$,$i_\sc$,$\Omega_\sc$) set, we compute a new
    best fit using a Levenberg-Marquard minimization \citep{Press_etal_1992} and
    a three-body model, but with eleven free parameters: the center of mass
    velocity $\gamma$, and for each planet, the semi-amplitude of radial
    velocity $K$, the period $P$, eccentricity $e$, mean anomaly $M$ and
    periastron $\omega$, all given at the initial epoch.
    Throughout the paper the initial conditions are given at the same initial
    epoch $T_0=2\,453\,000\ \mathrm{JD}$.

    At this point we have a complete description of the system with the mass of 
    the hosting star and 14 parameters (7 for each planets) as follows:
    \begin{itemize}
	\item 4 chosen : $i_\sigma$ and $\Omega_\sigma$
	\item and 10 fitted : $K_\sigma$, $P_\sigma$, $e_\sigma$, $M_\sigma$,
	and $\omega_\sigma$.
    \end{itemize}

    However, masses and elliptic coordinates are easier to manipulate for a
    dynamical study.
    Using Eq. \ref{eq:m1}, we obtain a system of two equations with two unknowns
    ($m_\sb$ and $m_\sc$) which is easily solved with a Newton algorithm.
    The semi-major axis are then obtained using Kepler's third law.


  \subsection{Numerical integrations \label{sec:numinteg}}
    For the numerical integrations we use the Newtonian equations with secular
    corrections for the relativity.
    The Newtonian part of the integration is carried out by the symplectic
    integrator SABAC4 of \citet{Laskar_Robutel_2001} with a step size of $0.02$ 
    year.
    The secular corrections for the relativity are computed from the
    perturbation formulae given in \cite{Lestrade_Bretagnon_1982}:
         
	\begin{align}
	    \label{eq:lestrade}
	    \Frac{\mathrm{d}a}{\mathrm{d}t} &= \mu\frac{2na^2e\sin{v}}{r^2\eta}
	\left[10\frac{a}{r} - 3\right] \ , \notag \\
	    \Frac{\mathrm{d}\lambda}{\mathrm{d}t} &=
	n + \mu \Frac{na}{r^2} \left\{8\Frac{a\eta^2}{r}+6\Frac{r}{a}-20\right.\notag\\
	&\phantom{= n + \mu \Frac{na}{r^2} \lbrace \  } \left.
	-\Frac{\eta^2-\eta}{e^2} \left[ -10\Frac{a^2\eta^2}{r^2} -\Frac{a\eta^2}{r}
	+ 18\Frac{a}{r}-7 \right]\right\} \ , \notag \\
	    \Frac{\mathrm{d}\varpi}{\mathrm{d}t} &=
	\mu \Frac{na\eta}{r^2e^2} \left[ -10\Frac{na\eta\sin{v}}{r^2}
	-\Frac{a\eta^2}{r} + 18\Frac{a}{r}-7 \right] \ , \notag \\
	    \Frac{\mathrm{d}e}{\mathrm{d}t} &= \mu \Frac{na\eta\sin{v}}{r^2}
	\left[ 10 \Frac{a}{r} - 7 \right] \ .
	\end{align}
    
    where $\mu = Gm_\ast/c^2$, $n$ is the mean motion, and $\eta = \sqrt{1-e^2}$.
    These equations are averaged to obtain the following first order secular
    perturbations:
      
	\begin{gather}
	    \label{eq:averaged}     
	     <\dot{a}> = 0 \ , \notag \\
	     <\dot{e}> = 0 \ , \notag \\
	     \quad <\dot{M}> =
	    \Frac{\mu n}{a} \left[ 6-\Frac{15}{\sqrt{1-e^2}} \right]\ ,\notag \\
	     \quad <\dot{\varpi}> = \Frac{3\mu n}{a(1-e^2)} \ .
	\end{gather}
    
    These corrections are computed every 100 steps, that is every two years,
    with the current values at the given step for $ e $, $ a $ and $ n $, for 
    each planet.
    These approximated equations have been successfully tested by comparison with INPOP
    \citep{Fienga_etal_2008}.



  \subsection{Stability threshold \label{sec:stabmap}}
    In order to study the stability of a given orbit, we use Laskar's frequency map
    analysis \citep[][]{Laskar_1990, Laskar_1993}.
    Using a numerical integration of the orbit over a time interval of length
    $T$, we compute a refined determination (in $\degr/yr$) of the mean motions
    $n_1$, $n_1^\prime$ obtained over two consecutive time intervals of length
    $T_1 = T/2$.
    The stability index $D_1 = | n_1^\prime - n_1|$ provides a measure of the
    chaotic diffusion of the trajectory.
    Small values close to zero correspond to a regular solution while high
    values are synonymous of strong chaotic motion \citep[][]{Laskar_1993}.

    In this paper we look at many different orbits for many different initial
    conditions to detect stable regions. This calls for a way to automatically
    calibrate a threshold for stability $D_\mathrm{lim}$.
    To that end we use a second stability index $D_2$.
    Using the same numerical integration, we compute two new determinations of 
    the mean motion $n_2$ and $n_2^\prime$ over two consecutive time intervals
    of length $T_2 = T_1/k$, where $k > 1$.
    In the case of quasi-periodic motion, the diffusion should be close to zero
    but it is limited by the precision of the determination of the frequencies.
    Since $D_1$ is computed over longer time intervals, the frequencies are
    better determined, and thus, $D_1$ should be, on average, smaller than
    $D_2$.
    On the contrary, for chaotic trajectories, the diffusion will
    increase on average for longer time intervals. We can then determine an approximated
    value $D_\mathrm{lim}$ for which:

	\be
	    \label{eq:dlim}
	    \left\{\begin{aligned}
		D_1 \gg D_\mathrm{lim} &\Longrightarrow D_1 > D_2 \ ,\\
		D_1 \ll D_\mathrm{lim} &\Longrightarrow D_1 < D_2 \ .\\
	       \end{aligned}
	    \right.
	\ee
    We will then be assured that the first kind of orbits is in general stable, while the
    latter is considered chaotic.

    This approach is best used statistically over a grid of initial conditions,
    especially when we try to use a small integration time.
    In order to determine $D_\mathrm{lim}$ for a particular diffusion grid, we
    look at the distribution of $D_1 < D_2$ trajectories as a function of $D_1$.
    We actually work with smoothed values $D_1^\mathrm{s}$ and $D_2^\mathrm{s}$ 
    of $D_1$ and $D_2$ to reduce the influence of the chaotic orbits whose
    mean motion diffusions are small by mere chance.
    They appear as low diffusion orbits inside high diffusion region.
    The smoothing function is a simple geometric mean over the closest
    neighbors.
    Other functions, such as a convolution with 2D Gaussian, have been tested,
    but do not yield significantly better results.

    We bin the $\log{D_1^\mathrm{s}}$ data in $0.5$ wide bins, and compute for
    each bin the percentage of $D_1^\mathrm{s} < D_2^\mathrm{s}$ orbits.
    Fig.~\ref{fig:bin} shows a typical distribution obtained from a diffusion
    grid (in this case the top panel of figure \ref{fig:aw_ae_90_cop}).
    It reproduces the behavior expected from Eq. \ref{eq:dlim}: low diffusion
    orbits tend, in great majority, to have their diffusion index diminish when 
    time increases.
        
        \bin
        
    We choose to define $D_\mathrm{lim}$ as the $D_1$ value for which $99\%$ of
    the trajectories exhibit $D_1 < D_2$.
    Graphically, it is the abscissa for which the curve in Fig.~\ref{fig:bin}
    crosses $y = 0.99$.
    In this example we get $\log{D_\mathrm{lim}} \approx -2.87$.
     
        \testdlim
     
    The $99\%$ threshold is actually a compromise that works in the majority of
    encountered cases: it minimizes the number of orbits wrongly flagged as
    stable (false stable) or unstable (false unstable). This is also 
    approximately the value for which the number of false stables and false
    unstables is equivalent. For percentages lower than $95\%$ the number of
    false unstables is nearly null. It is easy to understand since a higher
    percentage threshold implies a lower value for $D_\mathrm{lim}$, which in
    turns leads to very few actually unstable orbits flagged, while we might miss
    several stable ones, and vice-versa.
    To estimate the number of false stables and unstables, we recomputed the
    diffusion grid on a longer integration time, $2 \times 40000$ years, and use
    this grid as a reference (Fig.~\ref{fig:testdlim}).
     

\section{\label{sec:cop_edge}Review of the coplanar edge-on case}
  \subsection{Global dynamics}
    Following \citet{Correia_etal_2005}, we will study in more details the
    dynamics in the neighborhood of the 3-body fit obtained in the case of
    coplanar orbits with $\sin i_\sb = \sin i_\sc = 1$ (that is, the system seen
    edge-on).
    The best fit to the radial velocity data for this particular configuration is
    given in table \ref{tab:best_fit_cop_90}. It is different from the solution
    \textbf{S4} in \citet{Correia_etal_2005} (Table 4) as explained in
    Sect.~\ref{sec:data}. For the dynamical aspect of the system, the important
    change is the decrease in planet c's eccentricity. As a consequence regions
    outside resonances are expected to be more stable, and the environment of the
    fit should be less chaotic. However this new solution is still unstable: the
    outer planet is lost shortly after about 150 millions years.

    We look for possible nearby stable zones, keeping HD202206\,b parameters constant
    since they are much better constrained, with small standard errors. We assume
    for now that the system is coplanar and seen edge-on, that is with both
    inclination at $90\degr$, and $\Omega_\sc = 0\degr$. We let $a_\sc$,
    $e_\sc$, and $\omega_\sc$ vary. We always keep $M_\sc + \omega_\sc$ constant as it
    is much better constrained by the radial velocity data. This implies that when
    we change $\omega_\sc$, the mean anomaly $M_\sc$ varies accordingly. In the
    particular case where $\Omega_\sc = 0\degr$, this means that the initial mean
    longitude $\lambda_\sc$ is kept constant.
    For each initial conditions we compute the diffusion index $\log{D_1}$, and the
    square root of the reduced $\chi^2$.

        \bestfitcop

    Fig.~\ref{fig:aw_90_cop_large} shows a global picture of the dynamics around
    the fit, in the planes $(a_\sc,e_\sc)$ and $(a_\sc,\omega_\sc)$ of initial
    conditions. The step-sizes for $a_\sc$, $\omega_\sc$, $e_\sc$ are respectively
    $0.005 AU$, $2\degr$ and $0.004$. The other parameters were kept constant and
    taken from the fit \textbf{S1} (Table \ref{tab:best_fit_cop_90}).
    The level curves give the $\rsc$ value computed for each set of initial
    conditions. The
        color
    scale gives the diffusion index $\log{D_1}$. The
        yellow to red
    areas are very chaotic, mainly due to the large eccentricities and
    masses of both planets.

    The orbital solution \textbf{S1} lies inside the $\rsc = 1.5$ level curves, at
    the coordinates marked by a cross, inside a high diffusion
        (green)
    area. Several low diffusion
        (blue and dark blue)
    zones exist for which the orbits are stabilized either by mean motion
    resonances, or by locking of $\Delta\varpi$ around $0\degr$.
    Orbits stabilized by the corotation of the apsidal lines are the blue to
    black zones around $\omega_\sc = 190\degr$ (bottom panel).
    The width (in the $\omega_\sc$ direction) increases with the semi-major axis $a_\sc$,
    from $90$ degrees at $2$ AU, to nearly 360 degrees at 4 AU, since a wider libration of
    $\Delta\varpi$ around $0\degr$ is possible without close encounters when the distance
    between the two planets increases.
    The red to green more or less vertical stripes cutting through these zones mark
    mean motion resonances, which for the most part have a destabilizing effect in the two
    plans considered in Fig.~\ref{fig:aw_90_cop_large}. However,
    the stronger ones also have stable orbits:
    \begin{itemize}
	\item the $1/4$ MMR at 2.2 AU with two stable islands around $\omega_\sc = 260\degr$
	    and $\omega_\sc = 50\degr$.
	\item the 5:1 MMR at 2.5 AU with a notable stable island around $\omega_\sc = 70\degr$
	    where the best fit is located.
	\item the $1/6$ MMR at 2.8 AU with a stable island around $\omega_\sc = 0\degr$ or high eccentricity.
    \end{itemize}

  \subsection{Stable fit}
    We now take a closer look at the 5:1 mean motion resonance island
    around $a_\sc=2.5$ AU and $\omega_\sc=50\degr$, where we believe the system is
    presently locked.
    Fig.~\ref{fig:aw_ae_90_cop} was constructed the same way as Fig.~\ref{fig:aw_90_cop_large},
    but the step-sizes are now $0.0025$ AU for $a_\sc$, $0.5\degr$ for $\omega_\sc$,
    and $0.002$ for $e_\sc$.

    For eccentricities higher than $0.2$, orbits are very chaotic
	(red dots),
    as the outer planet undergoes close encounters with the inner body.
    At lower eccentricities we notice some lower diffusion orbits for $\omega_\sc >
    120\degr$. Those orbits lie far outside the resonance, but may be stable
    because of the low eccentricity of the outer planet, and apsidal locking
    mechanism. They are however too far from the best fit ($\rsc > 1.5$) and are 
    less likely to be a good guess of the actual configuration of HD202206 system.

    A very noticeable feature of this resonant island appearing in both panel, is
    the existence of two distinct stable regions, separated by chaotic orbits
    inside the resonance itself. The two stable regions actually correspond to
    two different critical arguments: $\lambda_\sb - 5\lambda_\sc + 4\varpi_\sc$ in
    the structure on the rim, and $\lambda_\sb - 5\lambda_\sc + \varpi_\sb +
    3\varpi_\sc$ in the center.

    \awcoplarge

    The orbital solution \textbf{S1} (white cross) lies very close to the latter, in a chaotic
    region (green and yellow dots) between the two stable parts of the resonant island.
    In fact, one can pick stable orbits with a $\rsc$ not significantly worse than the best fit.
    For instance, the orbital solution \textbf{S2} given in table \ref{tab:90_stable}
    is stable and has a $\rsc$ of $1.4136$ (marked by a filled white circle).
    The orbital elements are the same as \textbf{S1}, except for $a_\sc$
    which was adjusted from 2.4832 AU to $2.49$ AU.

	\begin{table}[ht]
	\caption{\label{tab:90_stable} Stable orbital parameters \textbf{S2} for the HD202206
	     system for $i_\sb=i_\sc=90\degr$ and $\Delta\Omega = 0\degr$.
	     Using the fit \textbf{S1} as a starting point (Table \ref{tab:best_fit_cop_90}),
	     we select a value for the semi-major axis of the outer
	     planet such that the system becomes stabilized in the 5:1 mean motion resonance.
	     This orbit is marked by a white filled circle in Fig.~\ref{fig:aw_ae_90_cop}.}
	\centering
	\begin{tabular}{llcc}
	\hline\hline
	  \textbf{Param.} & \textbf{S2} & \textbf{inner} & \textbf{outer} \\
	\hline
	  $a$ & [AU] & 0.8053 & \textbf{2.49} \\
	  $i$ & [deg] & 90 & 90 \\
	  $e$ & & 0.431 & 0.104 \\
	  $M+\omega$ & [deg] & 239.016 & 250.38 \\
	  $\omega$ & [deg] & 161.91 & 105.56 \\
	  $\Omega$ & [deg] & 0 & 0 \\
	  $m$ & [$\mbox{M}_{\mbox{\begin{tiny}Jup\end{tiny}}}$] & 16.59 & 2.179 \\
	\hline
	  Date & [JD-2400000] & \multicolumn{2}{c}{53000.00} \\
	  $\rsc$ & & \multicolumn{2}{c}{1.4136} \\
	\hline
	\end{tabular}
	\end{table}


  \subsection{Resonant and secular dynamics}
    The orbital solution \textbf{S2} was integrated over $5\;Gyr$.
    It remained stable, and displays a regular behavior during the whole time.
     
    Using frequency analysis on an integration over $1 Myr$, we determined its
    fundamental frequencies (table \ref{tab:freqF}).
    Following our notation, $n_\sb$ and $n_\sc$ are the mean motions.
    The secular frequencies are noted $g_1$ and $g_2$.
    Finally $l_\theta$ is the frequency associated with the resonance's critical
    angle $\theta$.
    The fundamental secular frequencies $g_1$ and $g_2$, related to the
    periastron of the inner and outer planet, correspond to the periods $P_1
    \approx 14.1\times 10^3 $ yr and $P_2 \approx - 791$ yr (the periastron of the outer
    planet is retrograde).

    Due to the mean motion resonance, a linear relation links the first four fundamental
    frequencies in Table \ref{tab:freqF}: $n_\sb - 5n_\sc + g_1 + 3g_2 = 0$.
    As a consequence, one of them is superfluous.
    A new fundamental frequency associated to the resonance $l_\theta$ replaces it.
         
	\begin{table}[ht]
	    \caption{\label{tab:freqF}Fundamental frequencies for \textbf{S2}. $n_\sb$ and
		$n_\sc$ are the mean motions, $g_1$ and $g_2$ are the secular frequencies
		associated with respectively $\varpi_\sb$ and $\varpi_\sc$, and $l_\theta$ is
		the libration frequency of the resonant angle $\theta = \lambda_\sb -
		5\lambda_\sc + \varpi_\sb + 3\varpi_\sc$.}
	    \centering
	    \begin{tabular}{crr}
	    \hline\hline
		& Frequency   & Period \\
		& $\degr/yr$ & $yr$ \\
	    \hline
		$n_\sb$ & 513.157691 & 0.70 \\
		$n_\sc$ & 102.363620 & 3.52 \\
		$g_1$ & 0.025537   & 14097.19 \\
		$g_2$ & -0.455042  & 791.13 \\
		$l_\theta$ & 4.492737 & 80.13 \\
	    \hline
	    \end{tabular}
	\end{table}
     
    The solution \textbf{S2} is trapped in the 5:1 mean motion resonance with the
    following main resonant argument:
	\begin{equation}
	    \theta=\lambda_\sb - 5\lambda_\sc + \varpi_\sb + 3\varpi_\sc \ .
	\end{equation}
    The variations of $\theta$ versus time are plotted in Fig.~\ref{fig:S90_thetaR_bw}
    (green line) and exhibit a libration around $\theta_0 = 0\degr$.
    We observe nonetheless that this libration is the modulation of several different
    terms with similar amplitudes of approximately $40\degr$, but on different time
    scales.
    This leads to a libration with an amplitude that can be higher than $180\degr$.
         
        \thetar
     
    In order to describe more accurately the behavior of $\theta$,
    we search for a quasi-periodic decomposition of $\theta(t)$.
    We start with a frequency decomposition using frequency analysis
    \citep{Laskar_2003} as:
	\be
	    \label{eq:decomp1}
	    \theta(t) = \sum_j{A_j\cos(\nu_j t + \phi_j)}
	\ee
    And then we decompose each frequency $\nu_j$ on the four fundamental frequencies
    ($n_c$, $g_1$, $g_2$, $l_\theta$):
	\be
	    \label{eq:decomp2}
	    \nu_j = \alpha_j n_\sc + \beta_j g_1 + \delta_j g_2 + \gamma_j l_\theta\ ,
	\ee
    where $\alpha_j,\beta_j,\delta_j$, and $\gamma_j$ are integers.
    Each term of the decomposition follows a D'Alembert-like relationship expressed
    in Eq.~\ref{eq:decomp3}.
	\be
	    \label{eq:decomp3}
	    \alpha_j+\beta_j+\delta_j=0
	\ee
    which can be used to simplify the expression of $\nu_j$. Indeed rearranging the right
    hand side of Eq.~\ref{eq:decomp2}, we can write:
	\be
	    \label{eq:decomp4}
	    \nu_j = \alpha_j n_\sc + \beta_j (g_1 - g_2) + (\beta_j+\delta_j)g_2 +
	    \gamma_j l_\theta\ ,
	\ee
    and using Eq.~\ref{eq:decomp3}, we get:
	\be
	    \label{eq:decomp5}
	    \nu_j = \alpha_j (n_\sc - g_2) + \beta_j (g_1 - g_2) + \gamma_j l_\theta\ .
	\ee
       
	\begin{table}[ht]
	\caption{Quasi-periodic decomposition of the resonant angle $\theta =
	\lambda_\sb - 5\lambda_\sc + \varpi_\sb + 3\varpi_\sc$ for an integration of the
	orbital solution \textbf{S2} over 1 million years.
	The decomposition is given in the form given by Eq.~\ref{eq:decomp1}, and each
	frequency $\nu_j$ is expressed as a combination of the fundamental frequencies
	in the form given by Eq.~\ref{eq:decomp5}.
	For the sake of brievity, we only give the first 30 terms.
	}
	\label{tab:theta}
	    \centering
	    \begin{tabular}{rrrrrrrrr}
	\hline
	\hline
	     j & $\nu_j$ , deg/yr & $A_j$ , deg & $\phi_j$ , deg & $\alpha_j$ & $\beta_j$ & $\gamma_j$ \\
	\hline
	 1 &     0.480579 &  53.892 &  26.688 &   0 &   1 &   0 \\ 
	 2 &     4.973316 &  36.043 & -132.474 &   0 &   1 &   1 \\ 
	 3 &     4.492737 &  34.786 & 111.145 &   0 &   0 &   1 \\ 
	 4 &     4.012157 &  21.409 &  -5.446 &   0 &  -1 &   1 \\ 
	 5 &     0.961158 &  21.399 & -36.372 &   0 &   2 &   0 \\ 
	 6 &  -512.651573 &  17.254 & 127.883 &  -5 &   3 &   0 \\ 
	 7 &  -513.612733 &  16.069 & -104.463 &  -5 &   1 &   0 \\ 
	 8 &     5.934475 &  16.268 & -79.250 &   0 &   3 &   1 \\ 
	 9 &     6.415054 &  16.795 &  37.274 &   0 &   4 &   1 \\ 
	10 &  -522.430683 &  15.118 &  85.394 &  -5 &  -8 &  -1 \\ 
	11 &  -512.170994 &  12.658 &  64.316 &  -5 &   4 &   0 \\ 
	12 &     5.453895 &  12.716 & -16.043 &   0 &   2 &   1 \\ 
	13 &  -513.132153 &  12.432 &  11.782 &  -5 &   2 &   0 \\ 
	14 &  -503.833622 &  11.911 & 117.837 &  -5 &  12 &   1 \\ 
	15 &     8.817951 &  11.908 &  79.987 &   0 &   9 &   1 \\ 
	16 &   205.156746 &  12.090 &  25.645 &   2 &  -1 &   0 \\ 
	17 &  -521.950104 &  10.953 &  21.911 &  -5 &  -7 &  -1 \\ 
	18 &  -522.911263 &  10.799 & 149.019 &  -5 &  -9 &  -1 \\ 
	19 &  -505.755940 &  11.882 & -168.212 &  -5 &   8 &   1 \\ 
	20 &  -504.314203 &   9.299 & -178.391 &  -5 &  11 &   1 \\ 
	21 &  -503.353041 &   8.914 &  53.782 &  -5 &  13 &   1 \\ 
	22 &  -505.275360 &  10.486 & 127.883 &  -5 &   9 &   1 \\ 
	23 &   102.818662 &  10.628 & 116.261 &   1 &   0 &   0 \\ 
	24 &  1018.888092 &   9.462 & -112.965 &  10 & -10 &  -1 \\ 
	25 &     6.895632 &  10.657 & -25.001 &   0 &   5 &   1 \\ 
	26 &   307.975409 &   8.924 & -128.711 &   3 &  -1 &   0 \\ 
	27 &     3.531578 &   8.135 & -122.054 &   0 &  -2 &   1 \\ 
	28 &     7.376215 &   8.880 &  88.862 &   0 &   6 &   1 \\ 
	29 &  1025.783728 &   7.756 & -49.947 &  10 &  -5 &   0 \\ 
	30 &  -411.274650 &   7.376 & 165.597 &  -4 &   0 &   0 \\ 
	\hline
	\end{tabular}
	\end{table}
     
    The first thirty terms of the decomposition can be found in Table
    \ref{tab:theta}. The decrease of the amplitudes $A_j$ is slow,
    due to the strong perturbations.

    The first term ($j=1$) is responsible for the long term oscillations with frequency
    $g_1 - g_2$. The period corresponding to $g_1 - g_2$, associated to the angle
    $\Delta\varpi=\varpi_\sb - \varpi_\sc$, is $P_{\Delta\varpi} \approx 749.09$ years.
    It is worth mentioning that the angle $\Delta\varpi$ is not in libration in this system,
    as opposed to what has been observed in planetary systems locked in a $1/2$ mean motion 
    resonance, such as GJ876 \citep{Laughlin_Chambers_2001, Lee_Peale_2002, Ji_etal_2002, Correia_etal_2009b}
    or HD89830 \citep{Gozdziewski_Maciejewski_2001}, or in a $3/2$
    mean motion resonance such as HD45364 \citep[][]{Correia_etal_2009}.
    The second one ($j=2 $) introduces the short term oscillations of
    frequency $l_\theta$, corresponding to a period $P_{\theta} \approx 101.49$ years.
    More precisely the libration of $\theta$ is made of three different kinds of
    contributions:
    \begin{itemize}
        \item secular terms: $\alpha_j=\gamma_j=0$, hence of the form
            $\beta_j(g_1-g_2)$;
        \item resonant terms: $\alpha_j=0$ and $\gamma_j \ne 0$, of the form
            $\beta_j(g_1-g_2)+\gamma_j g_\theta$;
        \item and short period terms: $\alpha_j \ne 0$.
    \end{itemize}
    We plotted with a solid line in Fig.~\ref{fig:S90_thetaR_bw} the contribution
    of the term $j=2$, the first secular contribution.
      


  \subsection{Test particles}
    We test in this section the possibilities for a third body around HD202206. To that
    end we integrate the \textbf{S2} solution with added massless particles over
    $16000$ years. As explained in the previous sections, we compute two
    determinations $n$ and $n^\prime$ of each particles mean motion over two
    consecutive time intervals of $8000$ years. We then obtain a stability index
    $D = \log{|n-n^\prime|}$ for each particle, which we plotted with a color code
    in Fig.~\ref{fig:testpart}. 
    Both panels are $(a,e)$ grids of initial conditions of the test particles.
    We vary the semi-major axis from $0.05$ AU to $0.5$ AU with a $0.0025$ step size
    in the left panel, and from $0.5$ AU to $10$ AU with a $0.05$ step size
    in the right panel.
    We span eccentricities of the test particles from $0$ to $0.9$ with a $0.005$ step size.
    Since the particles mean motions are higher in the left panel,
    they were integrated with a time step of $10^{-3}$ years instead of $2\times 10^{-3}$.

    Due to the very large eccentricities of HD202206\,b and HD202206\,c, the
    dynamical environment between and around them is very unstable. As a result,
    we don't expect any viable planets with a semi-major axis between approximately
    $0.12$ AU and $6.5$ AU (red dots). Most of these particles were actually lost
    before the end of the integration, either through collision or by having their
    eccentricity increased higher than $1$.
    The same computation with particles of one Earth mass yields very similar results.
    Assuming that \textbf{S2} is a good representation of the HD202206 planetary
    system, we can use these results to put constraints on hypothetical and yet undetected
    additional companions.
    There is clearly two possible regions for new planets: either close to the star ($a<0.12$ AU),
    or outside HD202206\,c ($a>6.5$ AU).

    In the first case, any planet massive enough should already have been detected as many full
    period are available in the data.
    Assuming a low eccentricity for the hypothetical companion and a 6 m/s instrumental precision
    we find that planets bigger than 24 earth masses should have already been detected.
    A Neptune-sized planet can exists anywhere between 0.06 AU and 0.12 AU, and
    a 10 earth masses planet, anywhere between 0.02 AU and 0.12 AU.

    In the second case ($a>6.5$ AU) the period is greater than 16 years, meaning that we have only
    covered approximately half an orbit at best. However a planet massive enough would create
    at least a detectable trend in the data.
    At 6.5 AU we can rule out the existence of a planet with more than half a Jupiter mass.
    At 10 AU we can rule out a planet between 1 $M_{Jup}$ and 3 $M_{Jup}$, depending on the phase.
    We conclude that a yet undetected planet smaller than half a Jupiter mass could exist at
    semi-major axis greater than 6.5 AU.

        \testparts

  \subsection{Finding the center of libration inside a resonance through frequency analysis}
    \subsubsection{Center of libration}
    We will consider in this section the planar three-body problem, and more particularly,
    the planetary problem with a $p:p+1$ mean-motion resonance.
    The problem is to find the orbit center of libration starting from a quasi-periodic
    orbit in the resonance.

    In the restricted case this center of libration is a well defined periodic orbit,
    such as the Lagrangian points in the $1:1$ resonance.
    However in the general problem it is not so easy to define and to find this orbit.
    It has 4 degrees of freedom, and its quasi-periodic orbits live on 4-torus in the phase space.
    It should be noted that the problem can be restricted to 3 degrees of freedom,
    using the angular momentum reduction.
    Each dimension of the 4-torus is associated to one of the four fundamental frequencies
    $(f_0,f_1,f_2,f_3)$ of the orbit.
    Let's suppose that $f_0$ is the frequency of the resonant mode.
    The orbit equivalent to the center of libration is living on a 3-torus, depending only on
    $(f_1,f_2,f_3)$.
    In other words, it is a torus where the fourth dimension associated to the resonance has
    a null amplitude.
    Intuitively, we can represent it has the center of the 4-torus in the dimension associated
    with the resonance.
         
        \torus

    \subsubsection{Quasi-periodic decomposition}
    If this orbit was periodic we could use a simple Newton algorithm to find it (provided that we start within the convergence radius),
    as it is a fixed point of a Poincar{\'e} map. This method has been extensively used in numerical search of periodic orbit families of the
    three body problem \citep[see for instance][]{Henon_1974, Henon_1997}.

    We present here a new numerical method to find quasi-periodic center of libration, using the fact
    that we can get an accurate quasi-periodic decomposition of a numerically integrated
    quasi-periodic orbit with frequency map analysis \citep{Laskar_2003}. Let $\mathbf{X}(t)$ be a state vector
    of such an orbit. Using frequency analysis, we can obtain a quasi-periodic representation
    for any component $x$ of the vector $\mathbf{X}$:

	    \be
	    \label{eq:qpd}
	    x(t) = \sum_{(k)}{A_{(k)}\mathrm{E}^{i <k,f> t}}
	    \ee
    for each coordinate of the vector $\mathbf{X}$, with $k = (k_0, k_1, k_2, k_3) \in \mathbb{Z}^4$, 
    $f = (f_0, f_1, f_2, f_3)$, and $A_{(k)} \in \mathbb{C}$.
    We can separate these sums in two parts: $x(t) = u(t) + v(t)$ where $u$ does not depend on
    the resonant frequency $f_0$, and $v$ has all the terms depending on $f_0$.

	    \be
		\label{eq:uv}
		\left\{\begin{aligned}
		    u(t) &= \sum_{(k), k_0 = 0}{A_{(k)}\mathrm{E}^{i <k,f> t}}\ ,\\
		    v(t) &= \sum_{(k), k_0 \neq 0}{A_{(k)}\mathrm{E}^{i <k,f> t}}\ .
		   \end{aligned}
		\right.
	    \ee

    The quasi-periodic orbit described by $\mathbf{U}(t)$ is precisely living on a 3-torus which has the characteristics we are looking for.
    However it is probably not a solution of the equations of motion.
    But we can assume that it is close to one.
    Hence, we can use $\mathbf{U(0)}$ as a new initial condition, and obtain a new resonant quasi-periodic orbit $\mathbf{X}^\prime(t)$ with:

	    \be
	    \label{eq:ic}
	    \mathbf{X}^\prime(0) = \mathbf{U}(0) \ .
	    \ee

    The amplitude of the resonant terms in this new orbit will be smaller.
    In other words, it lives on 4-torus closer to the 3-torus of the center of libration.
    We can then iterate this procedure to suppress all the terms with $f_0$.

    \subsubsection{Application to HD202206}
    We present here its application to the HD202206 system, using our orbital solution
    \textbf{S2} as a starting resonant orbit.
    We work with the $\mathbf{X} = (a_\sb, \lambda_\sb, z_\sb, a_\sc,\lambda_\sc, z_\sc)$ state vector,
    where $z_\sb = e_\sb \mathrm{E}^{i\varpi_\sb}$ and $z_\sc = e_\sc \mathrm{E}^{i\varpi_\sc}$.
    Each step $p$ of the method is decomposed as follows:
    \begin{itemize}
	\item Numerical integration of $\mathbf{X}^{(p)}$,
	\item Determination of $(f_0,f_1,f_2,f_3)$ using frequency analysis,
	\item Quasi-periodic decomposition: $x^{(p)}(t) \approx u^{(p)}(t) + v^{(p)}(t)$,
	\item New initial conditions: $x^{(p+1)}(0) = x^{(p)}(0) - v^{(p)}(0)$
    \end{itemize}
    The initial conditions are not computed exactly following Eq.~\ref{eq:ic}.
    Since we work with a finite number of terms, and the amplitudes of the terms in $v(t)$ are supposed to become small, the error we make because we only take into account that a finite number of terms will be smaller.

    The convergence proved to be fast as we reduced the amplitude of the most important resonant terms by 2 orders of magnitude in just 4 steps (Fig.~\ref{fig:steps}).
    We show graphically the decreasing amplitude of libration at each step in
    Fig.~\ref{fig:torus_num} where we plot approximated sections of the successive
    torus projected in the $(a_c,\varpi_c)$ plane.
    	\tores1
    No resonant terms are found in the first 100 terms of the quasi-periodic decomposition of each variable at the last step (with the exception of $\lambda_b$).
    In fact, in half of the variables, there are no resonant terms left in the
    first 300 terms (Fig.~\ref{fig:steps}).

	%
	\steps

    Once we find an orbit with a zero amplitude libration mode to a good approximation,
    we also get the 3-Torus it is living on since its quasi-periodic decomposition gives
    us a parametrization of the torus.
    The three angular variables of the torus appears in the decomposition:
	\be
	  \left\{\begin{aligned}
	    \phi_1 &= f_1 t + \psi_1 \\
	    \phi_2 &= f_2 t + \psi_2 \\
	    \phi_3 &= f_3 t + \psi_3
	  \end{aligned}
	  \right.
	\ee
    where $\psi_1$, $\psi_2$, and $\psi_3$ are initial phases.
    Let $\xi_k = \mathrm{E}^{i\phi_k}$.
    We can rewrite Eq.~\ref{eq:qpd} to reveal the underlying torus:
	\be
	x(\phi_1,\phi_2,\phi_3) = \sum_{(k)}{A_{(k)}
				       \xi_1^{\alpha_{(k)}}
				       \xi_2^{\beta_{(k)}}
				       \xi_3^{\gamma_{(k)}}}
	\ee

    Of all the orbits living on this torus we can choose the closest to the
    radial velocity data.
    This can be done by minimizing the $\chi^2$ on the three initial phases $\psi_1$,
    $\psi_2$, and $\psi_3$.

    We note \textbf{S3} the orbit obtained that way at step 6.
    We give initial conditions for \textbf{S3} in table \ref{tab:s3}.
    This solution yields a square root of reduced $\chi^2$ equal to 1.55.
    If the system is locked in the 5:1 mean motion resonance,
    the libration mode is likely to be dampened through dissipative processes.
    In that regard, the solution \textbf{S2} is unlikely as it is on the edge
    of the resonant island, and exhibit high resonant mode amplitude
    ($40\degr$ for the resonant critical angle).
    We expect that the real solution will be closer to \textbf{S3}.

	\begin{table}[ht]
	\caption{Orbital parameters of an orbit close to the center of libration of
	the 5:1 mean motion resonance. This orbit was obtain from \textbf{S2}
	(Table \ref{tab:90_stable}) at the sixth iteration.
	We expect that the actual system will be closer to this orbit
	as dissipative processes will have dampened the resonant mode.
	}
	\label{tab:s3}
	\centering
	\begin{tabular}{llcc}
	   \hline\hline
	      \textbf{Param.} & \textbf{S3} & \textbf{inner} & \textbf{outer} \\
	   \hline
	      $a$ & [AU] & 0.8050 & 2.5113\\
	      $i$ & [deg] & 90 & 90 \\
	      $e$ & & 0.439 & 0.071 \\
	      $M+\omega$ & [deg] & 239.25 & 247.30 \\
	      $\omega$ & [deg] & 161.81 & 78.19 \\
	      $\Omega$ & [deg] & 0 & 0 \\
	      $m$ & [$\mbox{M}_{\mbox{\begin{tiny}Jup\end{tiny}}}$] & 16.59 & 2.179 \\
	   \hline
	      Date & [JD-2400000] & \multicolumn{2}{c}{53000.00} \\
	      $\rsc$ & & \multicolumn{2}{c}{1.553} \\
	   \hline
	\end{tabular}
	\end{table}

%
%

\section{\label{sec:cop}Coplanar orbits}
  \subsection{Stability and low $\chi^2$ orbits}
    In this section we investigate the system behavior when both planets remain in
    the same orbital plane ($ I = 0\degr $) and are prograde, but the inclination to the line of
    sight is lower than $90\degr$:

    \begin{itemize}
        \item $i_\sb=i_\sc=i \leq 90\degr \ ;$
        \item $\Omega_\sc = 0\degr  \ .$
    \end{itemize}
    For each inclination value, we compute a new best fit with a
    Levenberg-Marquard algorithm. The $\rsc$ value obtained at different
    inclinations is plotted in Fig.~\ref{fig:cop_best} (solid line).
    Interestingly enough, we obtain better fits for lower inclination down
    to $15\degr$.

    In this configuration, the masses of the two companions grow approximately in proportion with
    $1/\sin{i}$ when the inclination $i$ diminishes. There is thus very little changes between
    $90\degr$ and $50\degr$, as seen in Fig.~\ref{fig:i_ae_large} (two top leftmost panels).
    For lower inclinations, due to the increased masses, mutual perturbations become stronger, and less orbits are stable. However, amongst the lowest $\rsc$ orbits, the ones in the 5:1 mean motion resonance remain stable for inclinations up to $15\degr$.

    For $i=10\degr$, no stable orbits are left (bottom rightmost panel). This puts clear limits on the inclination of the system, and on the masses of the two companions : $m_\sb < 95.5 \mjup$ and $m_\sc < 12.5 \mjup$.

	\aelarge

    An interesting property showing in Fig.~\ref{fig:i_ae_large} is that the
    lowest $\chi^2$ orbits are always in the vicinity of the stable island of
    the 5:1 mean motion resonance, for all inclinations. We believe that it is a
    strong point supporting the hypothesis of a HD202206 system locked in this
    resonance. It appears however, that after $i=50\degr$, and for lower
    inclinations, the low $\chi^2$ region is slightly shifted towards lower $a_\sc$ values than the resonance island.

    In order to have a more precise picture, we span $i$ values from $10\degr$ to $90\degr$ with a step size of $5\degr$.
    For each inclination value, we start by computing a new best fit with the
    Levenberg-Marquard algorithm.
    $\rsc$ and the stability indexes $D_1$ and $D_2$ are computed in the
    $(a_\sc , e_\sc)$ plane of initial conditions, around the best fit. The size of each grid
    is $101 \times 161$ dots, with step sizes of respectively $0.0025$ AU and $0.002$.
    For each of these maps, we compute $D_{lim}$ (see section \ref{sec:stabmap}),
    and detect stable regions, and their relative positions to the observations in
    terms of $\rsc$.

    In order to get a synthetic vision of all this data, we can get an
    estimation of the lowest stable $\rsc$ for each inclination (Fig.~\ref{fig:cop_best}).
    We plot with respect to $i$ the lowest $\rsc$ of $D^\mathrm{s}_1 < D_\mathrm{lim}$ orbits 
    (broken curve), along with the best fit value (solid curve).

	\progradebest

    It mainly confirms that the distance between stable orbits, and low $\rsc$ orbits is really small from $90\degr$ to $50\degr$. Is is actually slightly better between $60\degr$ and $50\degr$.

	\progradepct

    To have a better idea of the correlation between low $\rsc$ orbits and stable orbits, we look
    at the percentage of stable orbits inside a given $\rsc$ level curves (Fig.~\ref{fig:cop_pct}).
    It gives a synthetic representation of the overlap between low $\rsc$ regions and stable
    regions.
    It is very clear that inclinations lower than $30\degr$, although possible, are very unlikely. It also appears that inclinations between $90\degr$ and $50\degr$ are the most probable.

    If we limit the acceptable inclinations to $30\degr \le i \le 90\degr$,
    we can derive limits for the masses of the inner and outer planets
    (assuming a coplanar prograde configuration): 
    \begin{itemize}
	\item $1 \le m_b / m_b^{(0)} \le 2$
	\item  $1 \le m_c / m_c^{(0)} \le 2$
    \end{itemize}

    It is interesting to estimate the time when the radial velocity data will
    allow to determine a more accurate estimation of the inclination $i$,
    and masses for the system.
    Assuming this model is close enough to the true system, we look at
    the differences of the radial velocities in the $i=90\degr$ case, and
    $i=50\degr$, or $i=30\degr$.
    With the hypothesis of an instrumental precision of 
	7 m/s,
    we find (Fig.~\ref{fig:Delta_vr}) that we will have to
    wait until about 2015 to separate between $i=90\degr$ and
    $i=30\degr$, which corresponds to approximately a factor 2 in the masses.
    For $i=50\degr$, five more years are needed for the differences between the
    two configurations to be greater than the measurements precision.

	\deltavr

  \subsection{Resonant and secular behavior}
    We end this study of the coplanar configurations with a quick look at the
    dependence of the resonant and secular dynamics on the inclination.
    For each inclination, we picked a stable
    orbit with a low $\rsc$ value, and plotted its fundamental frequencies
    $l_\theta$, $g_1$, and $g_2$ in Fig.~\ref{fig:freqF}.

	\freqf

    As expected from the perturbation theory, when the masses increase,
    the secular frequencies also increase (in absolute value). We verify that it
    follows a rule in $1/\sin^2{i}$ (solid curve).
    This is a consequence of the fact that the most important terms responsible
    for the secular dynamics are of order two of the masses. 

\section{\label{sec:mutual}Mutually inclined orbits}
    In this section we drop the coplanarity constraint.
    We allow the inclinations $i_\sb$ and $i_\sc$ to vary independently,
    and we also allow variations of $\Omega_\sc$, the longitude of ascending
    node of $c$.
    To span the possible values for $i_\sb$, $i_\sc$, and $\Omega_\sc$ in an
    efficient manner, we restrict ourselves to two mass ratios $k_b$
    (Eq.~\ref{eq:mfactor}) for planet $b$:
    \begin{itemize}
      \item $k_b = 1$ ($i_b = 90\degr$),
      \item $k_b = 2$ ($i_b = 30\degr$),
    \end{itemize}
    and three mass ratios $k_c$ for planet $c$:
    \begin{itemize}
      \item $k_c = 1$ ($i_c = 90\degr$),
      \item $k_c = 2$ ($i_c = 30\degr$ or $i_c = 150\degr$),
      \item $k_c \simeq 5$ ($i_c = 11\degr$ or $i_c = 169\degr$).
    \end{itemize}
    For each couple of inclinations $i_\sb$ and $i_\sc$, we let $\Delta\Omega$
    vary between $0\degr$ and $360\degr$ with a $10\degr$ step size,
    and for each triplet $(i_\sb,i_\sc,\Delta\Omega)$ we perform a fit with the
    Levenberg-Marquard minimization, and compute a diffusion grid in the
    $(a_\sc,e_\sc)$ plane of initial conditions around the fit.
    The step sizes of the grids are respectively 0.0025 AU and 0.004.
    We plotted in Fig.~\ref{fig:mutone}c and Fig.~\ref{fig:muttwo}c the mutual
    inclination $I$ as a function of $\Delta\Omega$ for reference.
    For each configuration we look at the proportion of stable orbits inside the
    $\rsc=1.5$ level curve (Fig~\ref{fig:mutone}b and
    \ref{fig:muttwo}b).
    We will assume that stable zones with $\rsc \le 1.5$ orbits harbor
    potential solutions of the system.
    Note that we keep the initial inclination $i_b$ in $]0;\pi/2]$ as
    opposed to $i_c$ (see section~\ref{sec:conv}).
    Also for any given values $\alpha \in ]0;\pi/2]$ and $\beta \in [0;2\pi[$,
    assuming all the other orbital elements identical, the configuration
    ($i_b=\pi/2$, $i_c=\alpha$, $\Omega_c=\beta$) is symmetric to the
    configuration ($i_b=\pi/2$, $i_c=\pi-\alpha$, $\Omega_c=-\beta$) with
    respect to the plane $(\overrightarrow{i},\overrightarrow{k})$ (see
    Fig.~\ref{fig:geometry}).
    Since this plane contains the line of sight, the two configurations are
    indistinguishable using radial velocity measurements.
    Hence for $k_b=1$ ($i_b = 90\degr$) we will not look at $i_c = 150\degr$ and
    $169\degr$.

  \subsection{$k_b=1$}
        \mutone
    For $k_b=1$ ($i_b = 90\degr$, Fig.~\ref{fig:mutone}) we find that significant stable zones inside
    $\chi^2 = 1.5$ exist mostly for aligned and anti-aligned ascending nodes
    (i.e. $\Delta\Omega \simeq 0\degr$ and $180\degr$ in Fig~\ref{fig:mutone}b).

    For $k_c = 1$ (red curves) the aligned configurations ($\Delta\Omega  =
    0\degr \pm 10\degr$) are coplanar and prograde ($I \simeq 0\degr$),
    and the anti-aligned configuration ($\Delta\Omega = 180\degr \pm 10\degr$)
    are coplanar and retrograde.
    Outside those two particular cases, we find no significant stable zones for
    $\rsc \le 1.5$. Indeed, while the resonant island is roughly
    centered on the lowest $\chi^2$ level curve,
    it is not stable outside the coplanar configurations.
    We find that there exists an extended zone of stability outside the
    resonance for $\Delta\Omega \simeq 30\degr$, but it lies just outside the
    $1.5$ level curve.
    Note that due to symmetries, the situation for $\Delta\Omega > 180\degr$
    mirrors that of $\Delta\Omega < 180\degr$.

    For $k_c = 2$ (green curves) we find again stable zones for
    $\Delta\Omega \simeq 0\degr$, but not around $180\degr$.
    However stable regions with potential solutions exist for $\Delta\Omega$
    values up to $90\degr$, corresponding to mutual inclinations between
    $60\degr$ and $90\degr$.
    This is mostly due to the minimum $\chi^2$ getting smaller up to
    $\Delta\Omega = 90\degr$ (see the green curve in Fig.~\ref{fig:mutone}a).
    While the stable regions, both in the resonance and outside, shrink when
    $\Delta\Omega$ augments, the $\chi^2 = 1.5$ level curves encompass a larger
    area.

    Finally for $k_c = 5$ (blue curves), no significant stable zones are found
    at low $\chi^2$ values.

    To summarize, potential solutions (stable orbit with a low $\chi^2$) for
    $i_b = 90\degr$ mainly exists for coplanar configurations, inside the 5:1
    mean motion resonance. If planet c's mass is kept low ($k_b < 5$) stable
    regions do exist for non coplanar configurations, but they are located
    outside the lowest $\chi^2$ region. A noteworthy exception is the retrograde
    configuration, where stable orbit are only found close to the coplanar case,
    which only happens for $k_c = 1$ and $\Delta\Omega \simeq 180\degr$.
    
  \subsection{$k_b=2$}
    When we double the mass of planet $b$ ($i_b = 30\degr$,
    Fig.\ref{fig:muttwo}), once again retrograde potential solutions can only
    occur for coplanar orbits:
    $i_c = 150\degr$, and $\Delta\Omega \simeq 180\degr$ (green dotted curve).
    Concerning prograde orbits, there exists potential stable solutions for a
    mutual inclination of $I \simeq 60\degr$:
    \begin{itemize}
    	\item $i_c = 90\degr$ and $\Delta\Omega = 0\degr$ (red curve),
	\item $i_c = 30\degr$ and $\Delta\Omega = 180\degr$ (green solid
	curve).
    \end{itemize}
        \muttwo

  \subsection{Conclusions}
    To sum up we find that the configurations which have a significant stable
    zone at low $\chi^2$ values are found mostly when the two lines of nodes are
    aligned.
    That is for $\Delta \Omega$ close to $0\degr$ or $180\degr$.
    In addition, stable orbits with the lowest $\chi^2$ are all in the 5:1
    mean motion resonance except for nearly coplanar retrograde configurations,
    where they can also be close to the commensurability.
    However we can also find stable orbits outside the resonance in the prograde
    resonance, usually with $\rsc$ higher than 1.45.
    Retrograde configurations seem to be limited to nearly coplanar orbits with
    anti-aligned ascending nodes.
    Other than that, we could not find any clear correlation with mutual
    inclination.

    The fact that we find non-resonant stable solutions for
    retrograde configurations is consistent with \citet{Smith_Lissauer_2009}.
    They have shown that retrograde configurations allow more closely
    packed systems than prograde configurations.
    It was also suggested by \citet{Gayon_Bois_2008} that retrograde
    configurations are likely alternatives both from the radial velocity data
    and the long term stability point of view. 
    Forming such a system remains however difficult, and we will thus not favor
    this hypothesis.
     

\section{Discussion and conclusion}

    Assuming that the system is coplanar, we performed a systematic study of the dynamics of
    the system for different inclinations to the line of sight.
    We are able to find constraints for the inclination to the line of sight:
    $30\degr \le i \le 90\degr$.
    This means that the companions' masses are most likely not greater than
    twice their minimum values:
    \begin{itemize}
      \item $1 \le m_b / m_b^{(0)} \le 2$
      \item $1 \le m_c /m_c^{(0)} \le 2$
    \end{itemize}
    We also studied the influence of mutual inclination for two different inclinations
    of the planet $b$ ($i_b = 90\degr$ and $i_b = 30\degr$), but did not find any clear
    correlation other than that retrograde potential stable solution consistent with the
    radial velocity data seem to be limited to mutual inclinations close to $180\degr$
    (i.e. nearly coplanar orbits).
    As \citet{Godziewski_etal_2006}, we find possible stable solutions
    with low $\chi^2$ for a wide range of mutual inclinations between $0\degr$ and
    $90\degr$.
    The current data cannot yield more precise constraints.
    Also the masses determination is dependent on the stellar mass which is not well established.

    Although all published dynamical studies of HD202206 suggest that $b$ and
    $c$ are in a 5:1 mean motion resonance, it is still a debated question.
    For instance, \citet{Libert_Henrard_2007} assume that it is just close to the
    commensurability.
    Libration of $\Delta\varpi$ occurs for particular initial values of
    this angle, providing a stabilizing mechanism outside the mean-motion resonance,
    not far from the best fits.
    In most cases other than retrograde coplanar configurations, those orbits in
    near-commensurability are worse solutions than the ones in resonance, but they could
    be more probable if the eccentricities are overestimated (especially for $b$).
    We find that all significant stable zones with the best O-C are in
    the 5:1 mean motion resonance.
    In fact the minimum $\chi^2$ is almost always in the resonance or very close to it,
    and stable orbits in the resonance can be found with $\chi^2$ not significantly higher
    than the best fit.
    In addition the O-C level curves tend to follow roughly the resonant island,
    even though the agreement is not as perfect as for the HD45364 system
    \citep{Correia_etal_2009}.
    This is an improvement from \citet{Correia_etal_2005}
    where the best fit lay outside the resonant island, and the $\chi^2$ had to be degraded
    to find a stable solution. 
    We thus believe that the resonant configuration is the most probable.
    We provide a stable solution (\textbf{S2}, Table \ref{tab:90_stable}) in the coplanar
    edge-on case. This solution shows a high amplitude resonant mode in the libration of the
    critical angle. We believe that this resonant mode is probably dampened by dissipative
    processes.
    We use frequency analysis to find a tore on which such orbits exist.
    Although the specific orbit we give in Table \ref{tab:s3} does not have a
    very low $\rsc$ at 1.55, we expect that the true orbit will be close to it
    with a low libration amplitude.

    Note that for retrograde configurations, the picture is quite different.
    The best fit lies in a very stable region just outside the mean motion
    resonance.
    While these orbits are valid candidates from the dynamical and the
    observational points of view, we do not favor them as the formation of these
    systems is hard to explain.

    We investigated the possibility of undetected companions. We found that planets with masses
    smaller than approximately one Neptune mass can exist for semi-major axis lower than 0.12 AU.
    $0.5\,M_{Jup}$ planets are also possible beyond 6.5 AU.
    No planets are possible between 0.12 AU and 6.5 AU as they would be unstable.
    The two planets model may prove to be wrong in the future, but these hypothetical
    new companions should not have a big impact on the already detected ones.


\begin{acknowledgements}
	We acknowledge support from the Swiss National Research Found (FNRS),
	French PNP-CNRS, and Genci/CINES.
\end{acknowledgements}

\bibliographystyle{aa}
\bibliography{defs.bib,nemo.bib}

\newcommand{\noopsort}[1]{}
\begin{thebibliography}{23}
\expandafter\ifx\csname natexlab\endcsname\relax\def\natexlab#1{#1}\fi

\bibitem[{Correia {et~al.}(2009{\natexlab{a}})Correia, Couetdic, Laskar,
  Bonfils, Mayor, Bertaux, Bouchy, Delfosse, Forveille, Lovis, Pepe, Perrier,
  Queloz, \& Udry}]{Correia_etal_2009b}
Correia, A. C.~M., Couetdic, J., Laskar, J., {et~al.} 2009{\natexlab{a}},
  {\aap}, accepted

\bibitem[{Correia {et~al.}(2009{\natexlab{b}})Correia, Udry, Mayor, Benz,
  Bertaux, Bouchy, Laskar, Lovis, Mordasini, Pepe, \&
  Queloz}]{Correia_etal_2009}
Correia, A. C.~M., Udry, S., Mayor, M., {et~al.} 2009{\natexlab{b}}, {\aap},
  496, 521

\bibitem[{{Correia} {et~al.}(2005){Correia}, {Udry}, {Mayor}, {Laskar}, {Naef},
  {Pepe}, {Queloz}, \& {Santos}}]{Correia_etal_2005}
{Correia}, A.~C.~M., {Udry}, S., {Mayor}, M., {et~al.} 2005, {\aap}, 440, 751

\bibitem[{{Fienga} {et~al.}(2008){Fienga}, {Manche}, {Laskar}, \&
  {Gastineau}}]{Fienga_etal_2008}
{Fienga}, A., {Manche}, H., {Laskar}, J., \& {Gastineau}, M. 2008, \aap, 477,
  315

\bibitem[{Gayon \& Bois(2008)}]{Gayon_Bois_2008}
Gayon, J. \& Bois, E. 2008, {\aap}, 482, 665

\bibitem[{{Go{\'z}dziewski} {et~al.}(2006){Go{\'z}dziewski}, {Konacki}, \&
  {Maciejewski}}]{Godziewski_etal_2006}
{Go{\'z}dziewski}, K., {Konacki}, M., \& {Maciejewski}, A.~J. 2006, \apj, 645,
  688

\bibitem[{{Go{\'z}dziewski} \&
  {Maciejewski}(2001)}]{Gozdziewski_Maciejewski_2001}
{Go{\'z}dziewski}, K. \& {Maciejewski}, A.~J. 2001, \apjl, 563, L81

\bibitem[{Henon(1974)}]{Henon_1974}
Henon, M. 1974, Celestial Mechanics, 10, 375

\bibitem[{Henon(1997)}]{Henon_1997}
Henon, M. 1997, Generating Families in the Restricted {Three-Body} Problem, ed.
  M.~{Henon}

\bibitem[{{Ji} {et~al.}(2002){Ji}, {Li}, \& {Liu}}]{Ji_etal_2002}
{Ji}, J., {Li}, G., \& {Liu}, L. 2002, \apj, 572, 1041

\bibitem[{{Laskar}(1990)}]{Laskar_1990}
{Laskar}, J. 1990, Icarus, 88, 266

\bibitem[{{Laskar}(1993)}]{Laskar_1993}
{Laskar}, J. 1993, Phys. D, 67, 257

\bibitem[{{Laskar}(1999)}]{Laskar_1999}
{Laskar}, J. 1999, in NATO ASI Hamiltonian Systems with Three or more Degrees
  of Freedom, ed. C.~Simo (Kluwer), 134--150

\bibitem[{{Laskar}(2003)}]{Laskar_2003}
{Laskar}, J. 2003, ArXiv Mathematics e-prints

\bibitem[{{Laskar} \& {Robutel}(2001)}]{Laskar_Robutel_2001}
{Laskar}, J. \& {Robutel}, P. 2001, Celestial Mechanics and Dynamical
  Astronomy, 80, 39

\bibitem[{{Laughlin} \& {Chambers}(2001)}]{Laughlin_Chambers_2001}
{Laughlin}, G. \& {Chambers}, J.~E. 2001, \apjl, 551, L109

\bibitem[{Lee \& Peale(2002)}]{Lee_Peale_2002}
Lee, M.~H. \& Peale, S.~J. 2002, \apj, 567, 596

\bibitem[{{Lestrade} \& {Bretagnon}(1982)}]{Lestrade_Bretagnon_1982}
{Lestrade}, J.-F. \& {Bretagnon}, P. 1982, {\aap}, 105, 42

\bibitem[{{Libert} \& {Henrard}(2007)}]{Libert_Henrard_2007}
{Libert}, A.-S. \& {Henrard}, J. 2007, {\aap}, 461, 759

\bibitem[{{Press} {et~al.}(1992){Press}, {Teukolsky}, {Vetterling}, \&
  {Flannery}}]{Press_etal_1992}
{Press}, W.~H., {Teukolsky}, S.~A., {Vetterling}, W.~T., \& {Flannery}, B.~P.
  1992, {Numerical recipes in FORTRAN. The art of scientific computing}
  (Cambridge: University Press, 2nd ed.)

\bibitem[{Smith \& Lissauer(2009)}]{Smith_Lissauer_2009}
Smith, A.~W. \& Lissauer, J.~J. 2009, Icarus, 201, 381

\bibitem[{Sousa {et~al.}(2008)Sousa, Santos, Mayor, Udry, Casagrande,
  Israelian, Pepe, Queloz, \& Monteiro}]{Sousa_etal_2008}
Sousa, S.~G., Santos, N.~C., Mayor, M., {et~al.} 2008, {\aap}, 487, 373

\bibitem[{{Udry} {et~al.}(2002){Udry}, {Mayor}, {Naef}, {Pepe}, {Queloz},
  {Santos}, \& {Burnet}}]{Udry_etal_2002}
{Udry}, S., {Mayor}, M., {Naef}, D., {et~al.} 2002, {\aap}, 390, 267

\end{thebibliography}

\end{document}